\documentclass[12pt]{iopart}

\usepackage{bm}
\usepackage{graphicx}
\usepackage{amsfonts}
\usepackage{amsmath}
\usepackage{amssymb}
\usepackage{amsthm}
\usepackage{setspace}
\usepackage{helvet}
\usepackage{hyperref}
\newcommand{\bra}[1]{\langle #1 \bigr\rvert}
\newcommand{\ket}[1]{\bigl\lvert#1\rangle }
\newcommand{\braket}[2]{\langle #1\big|#2\rangle} 
\newcommand{\kla}[1]{\left( #1 \right)}
\newcommand{\klab}[1]{\left[ #1 \right]}

\begin{document}

\title{Hawking Radiation on an Ion Ring in the Quantum Regime}

\author{Birger Horstmann}
\address{Max-Planck-Institut f\"ur Quantenoptik, Hans-Kopfermann-Stra\ss e 1, 85748 Garching, Germany}
\author{Ralf Sch\"utzhold}
\address{Universit{\"a}t Duisburg-Essen, Lotharstra\ss e 1, 47048 Duisburg, Germany}
\author{Benni Reznik}
\address{Department of Physics and Astronomy, Tel Aviv University, Ramat Aviv 69978, Israel}
\author{Serena Fagnocchi}
\address{School of Physics and Astronomy, University of Nottingham, University Park, Nottingham NG7 2RD, United Kingdom}
\author{J. Ignacio Cirac}
\address{Max-Planck-Institut f\"ur Quantenoptik, Hans-Kopfermann-Stra\ss e 1, 85748 Garching, Germany}
\date{Garching, July 2010}

\begin{abstract}
This paper discusses a recent proposal for the simulation of acoustic black holes with ions \cite{Horstmann10}. The ions are rotating on a ring with an inhomogeneous, but stationary velocity profile. Phonons cannot leave a region, in which the ion velocity exceeds the group velocity of the phonons, as light cannot escape from a black hole. The system is described by a discrete field theory with a nonlinear dispersion relation. Hawking radiation is emitted by this acoustic black hole, generating entanglement between the inside and the outside of the black hole. We study schemes to detect the Hawking effect in this setup.
\end{abstract}

\pacs{04.70.Dy, 03.75.-b, 04.62.+v, 37.10.Ty}

\maketitle

\section{Introduction}
It was shown based on the theory of quantum fields (QFT) in curved spacetime that, surprisingly, black holes emit thermal radiation \cite{Hawking74}. However, the direct observation of Hawking radiation is difficult because the Hawking temperature is very small for astrophysical black holes. Furthermore, the original derivation of Hawking radiation relies on the validity of the wave equation on all scales, whereas QFT in curved space is expected to be reliable just up to the Planck scale. This paper is discussing Hawking radiation in the context of analogous hydrodynamical systems, which address these issues. These have an analog horizon at the transition from subsonic to supersonic (black hole) flow. It was shown that such analog horizons emit Hawking radiation \cite{Unruh81}, which can potentially be detected in experiments. The variety of different analog models and their tunability allows us to study the robustness of Hawking radiation against changes in the underlying microphysics. This will contribute to deepen our understanding of the Hawking effect also in gravitational black holes.


\begin{figure}[tbc]
\begin{center}
\includegraphics[width=150mm]{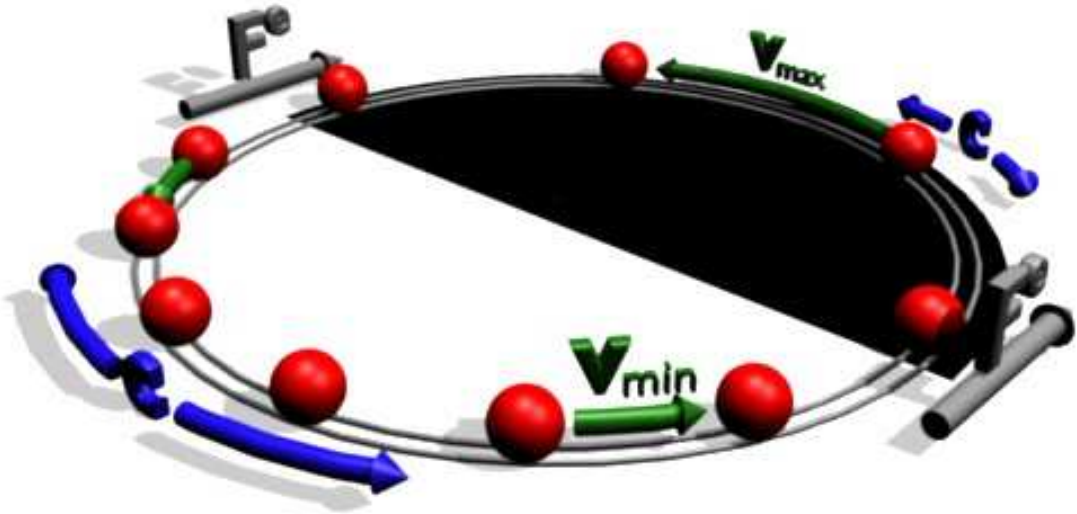}
\caption{Schematic depiction of the ion rotation with velocity $v\kla{\theta}$ and the phononic excitations with velocity $\pm c\kla{\theta}$, that depend on the varying ion spacing. In the subsonic region the ion velocity is $v\kla{\theta}=v_\text{min}$, in the supersonic region it is $v\kla{\theta}=v_\text{max}$. The external force $F^\text{e}\kla{\theta}$ localized at the transition between the super- and subsonic regions (de-) accelerates the ions. The black area represents the supersonic black hole region. A movie of the ion motion is accessible  \href{http://iopscience.iop.org/1367-2630/13/4/045008/media}{online} (New J. Phys. {\bf 13}, 045008 (2011)).}
\label{setup}
\end{center}
\end{figure}

In recent years many experimental tests of Hawking radiation have been proposed based on the hydrodynamical analogy \cite{Liberati05}, e.g., phonons in Bose-Einstein condensates (BECs) \cite{Cirac00,Carusotto08}, Fermi gases \cite{Giovanazzi05}, superfluid Helium \cite{Jacobsen98}, slow light \cite{Leonhardt00,Reznik00,Unruh03}, and nonlinear electromagnetic waveguides \cite{Schutzhold05}. Some experiments have implemented analog spacetimes in experiments, e.g., in BECs \cite{Lahav09}, optical fibres \cite{Philbin08}, and water surfaces \cite{Leonhardt08}. Common to all these attempts is that they have not reported on the observation of the quantum Hawking effect \footnote{After the submission of this paper the observation of the quantum Hawking effect has been claimed in \cite{Belgiorno10}. But pair emission of and entanglement between Hawking particles has not been observed yet.}.

In the present work we discuss a proposal how to build an analog model of a black hole in an experimentally realizable system of ions \cite{Horstmann10}. A special ingredient of our proposal is its discreteness, which naturally leads to a sublinear dispersion relation at high wavenumbers. This affects the trajectories of blue shifted waves close to the event horizon \cite{Jacobsen91}. The dispersion relation is, additionally, non-trivial at low wavenumbers because of the long range Coulomb force. However, as we will show, we still obtain Hawking radiation. Analytic derivations \cite{Jacobsen91,Jacobsen99,universality,group-phase} as well as numerical calculations show that the Hawking effect is robust against such short scale modifications, e.g., for a continuous field with a sublinear dispersion relation \cite{Unruh99} and a discretized field on a falling lattice \cite{Jacobsen99}. Our proposal uses a parameter regime which is accessible in experiments at temperatures currently achieved. Thus, it could lead to the experimental observation of Hawking radiation.

\begin{figure}[tbc]
\begin{center}
\includegraphics[width=80mm]{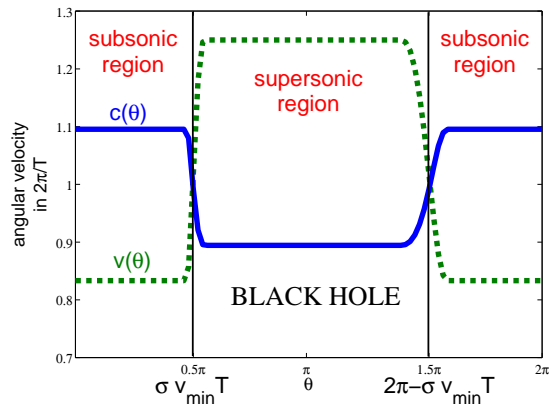}
\caption{Typical profiles of the ion ($v$, green) and the sound velocity ($c$, blue) used in this article (see \ref{section velocity profile}). The profile shows a super- and a subsonic region. The ion velocity in the subsonic region is $v_\text{min}=2\pi\times 0.8\bar{3}/T$ and the black hole horizon is located at $\sigma v_\text{min}T=2\pi\times 0.25$. A white hole horizon is also present on the ring \cite{Horstmann10}.}
\label{fig.velocity}
\end{center}
\end{figure}

Let us now summarize the main idea of our proposal. We are constructing a discrete analog of an hydrodynamical system with super- and subsonic regions in a quadrupole ring trap \cite{Drees64, Walther92} as schematically depicted in Fig. \ref{setup}. The ions are rotating on a ring with circumference $L$ with an inhomogeneous velocity profile $v\kla\theta$ (see Fig. \ref{fig.velocity}). Since the velocity profile should be stationary in the lab frame, the ions must be inhomogeneously spaced. Additional electrodes exerting a force $F^\text{e}\kla \theta$ on the ions generate the necessary (de-)acceleration. The oscillating displacements of the ions around this equilibrium motion are phonons with velocities $c\kla\theta\propto \kla{v\kla{\theta}}^{-1/2}$. Regions with sufficiently large ion spacings and sufficiently low phonon velocities are supersonic; here phonons can only move in the direction of the ion flow and are trapped like light inside a black hole. We consider a system with a super- and a subsonic region. The border between these regions is analogous to a black hole horizon and will be shown to emit Hawking radiation.

The plan of our paper is the following. First we will describe the system of ions on a ring and explain why it is expected to show the Hawking effect in Sec. \ref{system}. Then we briefly review previous works which are relevant for our analysis in Sec. \ref{Review}. After this preparation of the reader we present our simulation results and discuss them in Sec. \ref{simulations}. We are analyzing the Hawking effect in our system in two distinct ways: First we prove that the Hawking effect has a thermal spectrum by scattering pulses on the black hole horizon, second we are analyzing the emission of Hawking radiation from a black hole after its formation. In the latter situation we analyze the entanglement between pairs of Hawking particles, one inside and one outside the black hole. In this context we observe a transition from the \emph{quantum} to the \emph{classical} Hawking effect (spontaneous versus stimulated emission). Most importantly, this is suitable for experimental verification. In Sec. \ref{Experimental Parameters} we will discuss experimental setups which allow for the measurement of the analyzed physics.

\section{Ion Ring System}
\label{system}
In this section we are presenting our theoretical description of ions on a ring. In Sec. \ref{Section Hamiltonian} we start with the full Hamiltonian, explain our assumptions and approximations, and point to the equations underlying our simulations. We establish the connection with the general description of analog black holes in hydrodynamical system by considering the continuum limit of the system in Sec. \ref{section continuum limit}.

\subsection{Discrete Ion System}
\label{Section Hamiltonian}
In the following we are explaining the detailed setup of our proposal. The dynamics of $N$ ions with mass $m$ and charge $e$ are described by the Hamiltonian
\begin{equation}
\label{Hamiltonian general}
 \mathcal{H}=-\sum_{i=1}^{N} \frac{4\pi^2\hbar^2}{2mL^2}\frac{\partial^2}{\partial\theta_i^2}+\sum_{i=1}^{N}V^\text{e}\kla{\theta_i}+V^\text{c}\kla{\theta_1,\dots,\theta_N}
\end{equation}
with the Coulomb potential $V^\text{c}$ and a local external potential $V^\text{e}\kla{\theta}$. Instead of specifying $V^\text{e}$ we will impose an angular velocity profile $v\kla{\theta}$ by fixing the equilibrium positions. The required $V^\text{e}$ is then determined through the difference between the ion acceleration $\ddot{\theta}_i^0\kla t$ and the Coulomb force $F_i^\text{c}\kla{\theta^0_1,\dots,\theta^0_N}$. The Coulomb force and the external force are given in \ref{system detail}.

To this aim we impose the classical equilibrium positions
\begin{equation}
\label{definition g}
\theta_i^0(t)=g\kla{\frac{i}{N}+\frac{t}{T}},
\end{equation}
where $g$ maps the normalized indices $i/N\in [0,1]$ monotonically increasing onto the angles $\theta\in [0,2\pi]$ and is periodically continued. $g$ must be sufficiently smooth, i.e. three times continuously differentiable.
The stationary angular velocity profile is 
\begin{equation}
\label{stationary velocity}
v\kla{\theta}=\frac{g'\kla{g^{-1}\kla{\theta}}}{T}, 
\end{equation}
where $T$ denotes the rotation time of the ions (see Fig. \ref{fig.velocity}) and $g'$ is the derivative with respect to the argument of the function $g$. In a part of this paper we consider to dynamically create a black hole metric from a flat metric. To this aim, we decrease $v_\text{min}$ from the value $v_\text{min}=2\pi/T$ for homogeneously spaced ions in a Gaussian way with time constant $\tau$ (see \ref{section velocity profile}).

We choose a stationary velocity profile as depicted in \ref{fig.velocity}. It is composed of a subsonic region with the constant angular ion velocity $v\kla{\theta}=v_\text{min}$ in the angular range $0<\theta<\sigma v_\text{min}T$ and $2\pi-\sigma v_\text{min}T<\theta<2\pi$. In the supersonic region $\sigma v_\text{min}T<\theta<2\pi-\sigma v_\text{min}T$ the constant angular ion velocity is
\begin{equation}
\label{vmax}
v\kla{\theta}=v_\text{max}=\frac{2\pi}{T}\frac{1-\frac{2\sigma v_\text{min}T}{2\pi}}{1-2\sigma}.
\end{equation}
The two velocities $v_\text{min}$ and $v_\text{max}$ are naturally constrained by the rotation time of the ions $v_\text{min}<2\pi/T<v_\text{max}$. The black hole horizon in our system is located close to $\theta_\text{H}=\sigma v_\text{min}T$. Due to the ring structure a second horizon, the white hole horizon exists at $\theta_\text{H}=2\pi-\sigma v_\text{min}T$. The transitions at the horizons between the subsonic and the supersonic regions contain $2\gamma_1$ (black hole horizon) and $2\gamma_2$ ions (white hole horizon). The exact expression for the velocity profile is given in the Appendix in Eq. \eqref{metricg}.

We choose $v_\text{min}=2\pi\times 0.8\overline{3}/T$ and that the small transition regions $2\gamma_1,2\gamma_2$ contain $0.04N$ and $0.1N$ ions unless otherwise stated. If necessary, the black hole is dynamically created in the small time interval $\tau=0.05T$ (see Eq. \eqref{Reduce vmin}).

In this paper we do not work with the full Hamiltonian \eqref{Hamiltonian general}. Instead, we treat small perturbations around the equilibrium motion $\hat{\theta}_i\kla{t}=\theta_i^0\kla{t}+\delta\hat{\theta}_i\kla{t}$
and expand the Hamiltonian to second order in $\delta\hat{\theta}_i$
\begin{equation}
\label{Hamiltonian}
 \mathcal{H}=\frac{1}{2m}\sum_{i=1}^{N} \delta \hat{p}_i^2+\frac{m}{2}\sum_{i\ne j}f_{ij}(t){\delta\hat{\theta}}_i{\delta\hat{\theta}}_j
\end{equation}
with the time dependent force matrix $\mathcal{F}=\kla{f_{ij}}$ (see \ref{system detail}) and the canonical operators $\delta\hat{\theta}_i$ and $L\delta \hat{p}_i/(2\pi)=-i\hbar\partial_{\delta\theta_i}$ describing the phononic oscillations of the ions. This harmonic approximation is valid if the typical variation of the ion position is much smaller than the ion spacing
\begin{equation}
 \sqrt{\langle \delta\hat\theta^2 \rangle}  \approx \frac{2\pi}{L}\sqrt{\frac{\hbar}{m N\omega_\text{rot}}} \kla{\langle \hat{n} \rangle +\frac{1}{2} } \approx \frac{2\pi}{7\cdot 10^5} \kla{\langle \hat{n} \rangle +\frac{1}{2} } \ll \frac{2\pi}{N}.
\end{equation}
Here we use the typical mode frequency $N\omega_\text{rot}$ where $\omega_\text{rot}=2\pi/T$ is the rotation frequency of the ions. If we tune the ion velocity to lie in the same order of magnitude as the sound velocity, $\omega_\text{rot}$ gives the smallest mode frequency. We insert the experimental parameters considered in Sec. \ref{Experimental Parameters} with $N=1000$ here. In this paper we propose an experiment where the Hawking temperature $T_\text{H}$ and the initial temperature $T_0$ are not much larger than $\hbar\omega_\text{rot}/k_\text{B}$, so we can assume $\langle \hat{n}\rangle\sim 1$. Then the above requirement is nicely fullfilled.

The quasi-free quantum dynamics of this harmonic system \eqref{Hamiltonian} are governed by the classical linear equations of motion for the first and second moments. Through Wick's theorem all higher order correlation functions are determined by these moments. The first moments are $\langle \hat{\xi}_i\rangle$ with the definition
\begin{equation}
\label{first moments}
 \hat{\xi}_i=\begin{cases}
        \delta\hat{\theta}_i & i\in\{0,\dots,N-1\}\\
	-i\hbar\hat{\partial_{\theta_i}} & i\in\{N,\dots,2N-1\}
       \end{cases}.
\end{equation}
The second moments are grouped under the covariance matrix
\begin{equation}
\label{Gamma}
\Gamma_{ij}=\frac{1}{2\hbar}\langle\{\hat{\xi}_i\hat{\xi}_j\}_+\rangle,
\end{equation}
where $\{\}_+$ denotes an anticommutator. The equations governing the dynamics and determining the equilibrium states of the first and second moments are given in \ref{equations of motion}.

A stability analysis of this system is described in \ref{stability analysis}. We find that exponential instabilities, though present in this system, are not important for the proposed experiment because it is performed during only one rotation period $T$.

\subsection{Continuum Limit}
\label{section continuum limit}
\begin{figure}[tbc]
\begin{center}
\includegraphics[width=80mm,height=60mm,angle=0]{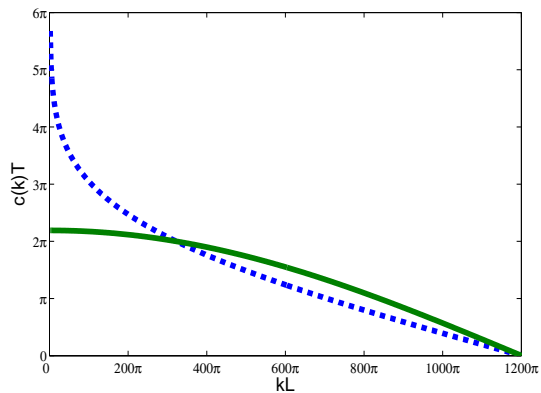}
\caption{Phononic group velocity $c(k)$ in the flat subsonic region as a function of $k$ for full Coulomb interactions (blue dashed line) and nearest-neighbor interactions only (green solid line). For nearest-neighbor interactions only the group velocity approaches a constant at small wavenumbers displaying the linear dispersion. For full Coulomb interactions a logarithmic divergence at small wavenumbers is observed for a finite system size. We use $\sigma v_\text{min}T=2\pi\cdot 0.375$, $N=1000$, and $e^2/4\pi\epsilon_0=1.2591/(2N)\cdot mL^3T^{-2}$ (see \ref{section velocity profile})}.
\label{dispersion}
\end{center}
\end{figure}
In order to get some insight, we consider the limit of an infinite number of ions and formulate the analogy with the standard Hawking effect in this limit.

We will first study the behavior of the dispersion relation of the ions at small wavenumbers (see Fig. \ref{dispersion} for the finite chain). We use the approximation given in \cite{Porras08} for small wavenumbers in the open Coulomb chain taking into account the long range Coulomb interactions. From
\begin{equation}
\omega(k)=ck\sqrt{1-\frac{2}{3}\log\kla{\frac{ka}{2}}}
\end{equation}
with the ion spacing $a=L/N$ we get
\begin{equation}
\label{phase group velocity}
\frac{d\omega(k)}{dk}=\frac{\omega}{k}-\frac{1}{3}\frac{c}{\sqrt{1-\frac{2}{3}\log\kla{\frac{ka}{2}}}}
\end{equation}
and
\begin{equation}
\frac{d^2\omega(k)}{dk^2}=\frac{-c}{3k\sqrt{1-\frac{2}{3}\log\kla{\frac{ka}{2}}}}\klab{1+\frac{1}{3\kla{1-\frac{2}{3}\log{\kla{\frac{ka}{2}}}}}}
\end{equation}
For $k=2\pi/L\cdot n$ with $1 \le n\le n_0\ll N$ this can be summarized as
\begin{equation}
\frac{d\omega(k)}{dk}\approx \frac{\omega}{k}\text{, and } \frac{d^2\omega(k)}{dk^2}\approx 0.
\end{equation}
For a given $k$, the phase and group velocity become identical in the continuum limit $N/L\to\infty$ as explicitly shown in Eq. \eqref{phase group velocity}. This is important for the definition of the Hawking temperature in the following, see Eq. \eqref{Hawking temperature}. If group and phase velocity did not coincide, it would not be clear a priori how to determine the correct Hawking temperature. For example, in the scenario considered in \cite{group-phase}, the product of group and phase velocity enters the formula for the Hawking temperature. Since both velocities are identical in the continuum limit for our proposal, for the finite system we can determine the Hawking temperature at a given $k$ from the group velocity.


Now, we calculate the Lagrangian for the scalar field $\hat{\Phi}\kla{\theta^0_i\kla{t},t}=\delta\hat{\theta}_i\kla{t}$ in the continuum limit. Here we can make an analogy with the standard Hawking effect as observed in \cite{Unruh81}. Because of the equilibrium motion of the ions the kinetic energy $K$ transforms according to
\begin{eqnarray}
K&=&\frac{m}{2}\kla{\frac{L}{2\pi}}^2\sum_{i=1}^N \kla{\frac{d\theta_i}{dt}}^2 \nonumber \\
&\approx& \int_0^{2\pi} d\theta \frac{\rho\kla\theta}{2}  \kla{\frac{d}{dt} \Phi\kla{\theta^0_i\kla{t},t}}^2 \nonumber \\
&=& \int_0^{2\pi} d\theta \frac{\rho\kla\theta}{2}  \kla{\partial_t\Phi+v\kla\theta\partial_\theta\Phi}^2,
\end{eqnarray}
where we introduced the conformal factor $\rho\kla\theta=n\kla\theta \cdot mL^2/(2\pi)^2$ with the density $n\kla{\theta}=N/(v\kla\theta T)$. For an homogeneous system the potential energy $V$ transforms to 
\begin{eqnarray}
V&=&\frac{m}{2}\sum_{i,j=1}^N f_{ij} \delta\theta_i\delta\theta_j=\sum_{\substack{k=0}}^{N-1}\frac{m}{2}D(k)^2|\Phi'_k|^2\nonumber\\
&=&\sum_{\substack{k=0}}^{N-1}\sum_{\substack{n,m=1\\\theta_n=\frac{2\pi}{L}n}}^N{\frac{L^2}{(2\pi)^2}}\frac{m}{2N}D(k)^2e^{-ik (\theta_n-\theta_m) }\Phi(\theta_n)\Phi(\theta_m)\nonumber\\
&\approx&\int_0^{2\pi}d\theta\frac{\rho\kla\theta}{2}\kla{iD(-i\partial_\theta)\Phi(\theta)}^2,
\end{eqnarray}
where $D\kla{\theta,k}=c\kla{\theta}k+\mathcal{O}\kla{k^3}$ is the dispersion relation of the Coulomb chain. Assuming a slowly varying $v\kla{\theta}$, we can now formulate the Lagrangian for the ion system in the laboratory frame in the continuum limit
\begin{equation}
\label{analog}
 \mathcal{L}=\int d\theta\frac{\rho\kla{\theta}}{2}\left[\kla{\partial_t\hat\Phi+v\kla\theta\partial_{\theta}\hat\Phi}^2-\kla{iD\kla{\theta,-i\partial_\theta}\hat\Phi}^2\right].
\end{equation}
This scalar field satisfying a linear dispersion relation at low wavenumbers with sound velocity
\begin{equation}
\label{eq.sound velocity}
\kla{c\kla{\theta}}^2=\frac{2\cdot n\kla{\theta}}{m} \kla{\frac{2\pi}{L}}^3 \frac{e^2}{4\pi\epsilon_0}
\end{equation}
for nearest-neighbor interactions, resulting from the actual form of the matrix $f_{ij}$ given in Eq. \eqref{eq.forcematrix}, is analogous to a massless scalar field in a black hole spacetime as first shown in \cite{Unruh81}. Its quanta cannot escape a supersonic region with $v\kla{\theta}>c\kla{\theta}$ like photons trapped inside a black hole. The horizon of this analog model is located at $c\kla{\theta_\text{H}}=v\kla{\theta_\text{H}}$ with $\theta_\text{H}\approx \sigma v_\text{min}T$. Following \cite{Unruh81}, pairs of Hawking particles are emitted close to the black hole horizon with a black body distribution at the Hawking temperature
\begin{equation}
\label{Hawking temperature}
 \frac{k_\text{B} T_\text{H}}{\hbar}=\frac{\kappa}{2\pi}=\frac{1}{4\pi v}\frac{d}{d\theta}\kla{v^2-c^2}|_\text{H}=\frac{3}{4\pi T}\frac{g''\kla{g^{-1}\kla{\theta}}}{g'\kla{g^{-1}\kla{\theta}}}\bigr|_{\theta=\theta_\text{H}}.
\end{equation}
The first equality defines the surface gravity $\kappa$ (see \cite{Hawking74,Unruh81}), in terms of the Hawking temperature. The second equality is derived in reference \cite{Unruh81}, the third one results from the explicit forms for $v\kla\theta$ and $c\kla\theta$ for nearest-neighbor interactions. 

\begin{figure}[tbc]
\begin{center}
\includegraphics[width=80mm,angle=0]{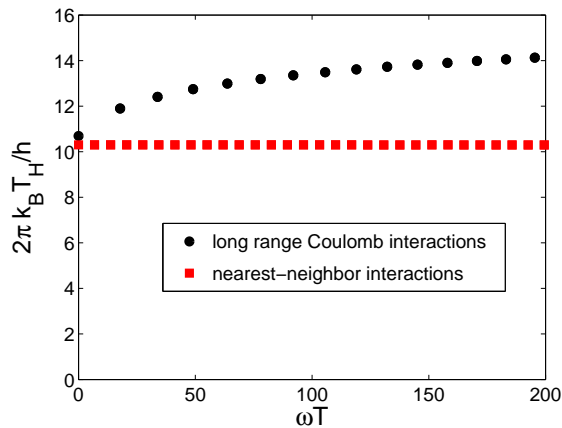}
\caption{Dependence of the Hawking temperature $T_\text{H}$ on the discrete mode frequencies $\omega(k)$ in the comoving frame at small frequencies. For nearest-neighbor interactions, the variation with the wavenumber is not visible, for long range Coulomb interactions variations are present. This is due to the nonlinear dispersion. Since these variations are not very large, we may approximately use a single Hawking temperature for the full system. We use $\sigma v_\text{min}T=0.375$, $N=1000$. (red) Nearest-neighbor interactions with $e^2/4\pi\epsilon_0=\frac{1}{2N}\frac{mL^3}{T^2}$; (black) Full Coulomb interactions with $e^2/4\pi\epsilon_0=\frac{1}{2N}\frac{mL^3}{T^2}$ (see \ref{section velocity profile}).}
\label{temperature}
\end{center}
\end{figure}

In the case of long-range Coulomb interactions, the calculation of the Hawking temperature $T_\text{H}$ (see Eq. \eqref{Hawking temperature}) is difficult because of the non-linear dispersion relation at low wavenumbers. In order to estimate the impact of this dispersion, let us introduce an effective wavenumber-dependent Hawking temperature. Because the ion spacings are inhomogeneous, we use a local density approximation. At each angle $\theta$ we calculate the local density $n(\theta)$ of the ions. We then extract a group velocity $c(\theta)$ from the analogous homogeneous system with constant density and $N_\theta=\left[n(\theta)\right]$ ions, where the square brackets denote rounding to the nearest integer. The group velocity is calculated from the dispersion relation at adjacent wavenumbers. With this method we calculate angle- and wavenumber-dependent group velocities $c(\theta,k)$. By comparing $c(\theta,k)$ with the ion velocities $v(\theta)$, we find for each wavenumber the black hole horizon $\theta_\text{H}$. The derivative of $c$ in Eq. \eqref{Hawking temperature} is performed with respect to the ion number $N_\theta$ instead of the angle $\theta$ so that only a single rounding procedure $N_\theta=\left[n(\theta)\right]$ is needed. Thus, Eq. \eqref{Hawking temperature} is transformed into 
\begin{align}
  \frac{k_\text{B} T_\text{H}\kla{k}}{\hbar}&=\frac{1}{2\pi T}\frac{g''\kla{g^{-1}\kla{\theta}}}{g'\kla{g^{-1}\kla{\theta}}}\biggr|_{\theta=\theta_\text{H}}\kla{1+\frac{N_{\theta_\text{H}}}{g'\kla{g^{-1}\kla{\theta}}^2}\biggr|_{\theta=\theta_\text{H}}\frac{\partial}{\partial N_\theta}c\kla{N_\theta,k}\biggr|_{N_\theta=N_{\theta_\text{H}}}}\nonumber\\
&\approx\frac{1}{2\pi T}\frac{g''\kla{g^{-1}\kla{\theta}}}{g'\kla{g^{-1}\kla{\theta}}} \kla{1+N_{\theta_\text{H}}\frac{c\kla{N_{\theta_\text{H}}+1,k}-c\kla{N_{\theta_\text{H}},k}}{g'\kla{g^{-1}\kla{\theta}}^2}}\biggr|_{\theta=\theta_\text{H}}.
\label{temperatureLDA}
\end{align}
The dependence of the Hawking temperature on the wavenumber is depicted in Fig. \ref{temperature} for the parameters used in the analysis in Sec. \ref{sec.scattering}. For nearest-neighbor interactions, the variation with the wavenumber is not visible, for long range Coulomb interactions variations are present. This is due to the nonlinear dispersion relation.                                                                                                                                                                                                                                                                                                                                                                                                                                                                                                                                                                                                                                                                                                                                                                                                                                                                                                                                                                                                                                                                                                                                                                                                                       We make an arbitrary choice and use one of these temperatures at small wavenumbers to analyze the whole system at all wavenumbers. In Sec. \ref{sec.scattering} we will discuss how our results depend on this choice. Note that in the continuum limit the dispersion relation for full Coulomb interactions becomes linear which resolves the arbitrariness. Thus we find a single Hawking temperature for the full system based on the form of the velocity profile. 

\section{Review}
\label{Review}
In this section we will give a short summary of previous works related to our proposal (see \cite{Balbinot2006} for further details). In Sec. \ref{sec.review.mode} we are describing a basic derivation of Hawking radiation with quantum field theory in curved spacetime for systems with a strictly linear dispersion relation. Secondly, we present a theoretical method that tests whether the expected radiation has a thermal spectrum \cite{Unruh99,Jacobsen99} for systems with sublinear dispersion relations (see Sec. \ref{Review Scattering}). Then we explain a proposal, originally aiming at Bose Einstein Condensates \cite{Carusotto08}, on how to detect Hawking radiation in an experiment as quantum correlations which emerge between a supersonic and a subsonic region, i.e. between the inside and the outside of a black hole (see Sec. \ref{Review Correlations}).

\subsection{Hawking radiation and Mode Conversion}
\label{sec.review.mode}
In this section we will summarize the derivation of Hawking radiation for a massless scalar field with a linear dispersion relation $D\kla{k}=ck$ in a black hole analog spacetime, defined by a velocity profile $v\kla\theta$ (see Eq. \eqref{vel} and Eq. \eqref{metricg}) and a density profile $\rho\kla{\theta}$. This system is governed by the Lagrangian
\begin{equation}
\label{analogconstantc}
 \mathcal{L}=\int d\theta\frac{\rho\kla{\theta}}{2}\left[\kla{\partial_t\hat\Phi+v\kla\theta\partial_{\theta}\hat\Phi}^2-c^2\partial_\theta\hat\Phi^2\right].
\end{equation}
We will in the following assume $\rho\kla\theta$ to be constant and work with the field $\Psi=\sqrt\rho\Phi$. The correspondend classical field equation is
\begin{equation}
\label{field equation}
 \left[\kla{\partial_t+\partial_\theta v\kla{t,\theta}}\kla{\partial_t+v\kla{t,\theta}\partial_\theta}-c ^2\partial_\theta^2\right]\Psi\kla{t,\theta}=0.
\end{equation}
The solutions of this equation determine the notion of excitation modes in the system and are of the form
\begin{equation}
 \Psi\kla{t,\theta}=\sum_\pm \int_0^\infty d\omega \kla{\Psi^\pm_\omega\kla{t,\theta}+H.c.}
\end{equation}
with $\Psi^\pm_\omega\kla{t,\theta}=\exp\kla{\pm i \omega t}\Psi_\omega\kla{\theta}$ for a stationary spacetime. $\omega$ denotes the frequency in the lab frame, it is related by $\omega=(v\pm c)k\cdot L/(2\pi)$ to the frequency $\pm ck$ in the comoving frame. The solutions $\Psi_\omega$ are normalized with respect to the Klein-Gordon inner product
\begin{equation}
\label{Klein-Gordon inner product}
 \braket{\Psi_\omega}{\Psi_{\omega'}}=
\frac{-i}{2}\int_0^{2\pi} d\theta \left[\Psi^*_\omega\kla{\partial_t+v\partial_{\theta}}\Psi_{\omega'}-\Psi_{\omega'}\kla{\partial_t+v\partial_\theta}\Psi^*_{\omega}\right].
\end{equation}
For each frequency $\omega$ we can find four independent modes $\Psi_\omega^\pm$/$\Psi_\omega^{\pm *}$. Complex conjugation relates modes with a positive/negative sign of the Klein-Gordon norm $\mathcal{N}_\omega=\braket{\Psi_\omega}{\Psi_{\omega}}$, corresponding to positive/negative frequency modes or particles/anti-particles. Note that the frequency in the comoving frame $\pm ck$ determines the sign of $\mathcal{N}$. The index $\pm$ denotes modes which are left-/rightmoving (up-/downstream) in the comoving frame. Finally, the scalar field theory is quantized by expanding the field operator in the modes (we drop the summation index $\pm$ now)
\begin{equation}
\hat{\Psi}=\sum_\omega\kla{\hat{a}_\omega \Psi_\omega+\hat{a}^\dagger_\omega \Psi^{*}_\omega}
\end{equation}
and postulating canonical bosonic commutation relations for the mode operators $\hat a_\omega$
\begin{equation}
 \left[\hat a_\omega,\hat a_{\omega'}\right]=0, \hspace{2mm}\left[\hat a^\dagger_\omega,\hat a_{\omega'}\right]=\delta_{\omega,\omega'}.
\end{equation}
The vacuum of the system is defined to be annihilated by all $\hat a_\omega$
\begin{equation}
 \hat{a}_\omega\ket{0}=0,\hspace{2mm}\forall\omega.
\end{equation}
We want to study the following time dependent situation: At initial times, called \emph{in}, the black hole is not present, the vacuum state is denoted $\ket{\text{in}}$ with corresponding modes $\hat a_\omega^\text{in}$. Then a black hole is created. At final times, called \emph{out}, the vacuum state of the system is $\ket{\text{out}}$ with corresponding modes $\hat a_\omega^\text{out}$. We are here interested in the time evolution of the system from initial to final times starting from the vacuum state of the system $\ket{\text{in}}$. The result of this time evolution can be described by a Bogoliubov transformation on the classical modes or the quantum mode operators
\begin{gather}
\label{Bogoliubov transformation}
\Psi_\omega^\text{out}=\sum_\omega \kla{\alpha_{\omega\omega'}\Psi_{\omega'}^\text{in}+\beta_{\omega\omega'}\Psi_{\omega'}^{in*}},\\
\hat{a}_\omega^\text{out}=\sum_\omega \kla{\alpha_{\omega{\omega'}}^{*}\hat{a}_{\omega'}^\text{in}-\beta_{\omega\omega'}^{*}\hat{a}_{\omega'}^{in\dagger}}
\end{gather}
with the Bogoliubov coefficients $\alpha_{\omega\omega'}$ and $\beta_{\omega\omega'}$. For a static and continuous system the frequency $\omega$ is conserved and the Bogoliubov coefficients are diagonal, e.g., $\beta_{\omega\omega'}=\delta_{\omega,\omega'}\beta_\omega$. The coefficients are related through $|\alpha_\omega|^2-|\beta_\omega|^2=1$. The particle content of the state $\ket{\text{out}}$ in terms of the state $\ket{\text{in}}$ is related to the Bogoliubov coefficients through
\begin{equation}
 \bra{\text{in}} \hat N_\omega^\text{out} \ket{\text{in}}=\sum_{\omega'}|\beta_{\omega\omega'}|^2=|\beta_\omega|^2.
\end{equation}
Therefore, it is nonvanishing if the time evolution from initial to final times mixes positive and negative frequency modes. In particular, the production of Hawking particles can be understood as the evolution of initial positive frequency modes into final negative frequency modes. The Bogoliubov coefficients of the inverse transformation have the same modulus as the ones of the forward transformation
\begin{gather}
\Psi_\omega^\text{in}=\sum_\omega \kla{\alpha^*_{\omega\omega'}\Psi_{\omega'}^\text{out}-\beta_{\omega\omega'}\Psi_{\omega'}^{\text{out}*}},\\
\hat{a}_\omega^\text{in}=\sum_\omega \kla{\alpha_{\omega{\omega'}}\hat{a}_{\omega'}^\text{out}-\beta_{\omega\omega'}^{*}\hat{a}_{\omega'}^{\text{out}\dagger}}.
\end{gather}
We will employ this relation to numerically determine the Bogoliubov coefficients for the discrete case of ions on a ring.

In the following we will briefly summarize the analytical calculation of the Bogoliubov coefficients for a continuous system with a strictly linear dispersion relation. One can decompose the field operator at late time into an upstream $\hat{\Psi}^-$ and a downstream $\hat{\Psi}^+$ part
\begin{equation}
 \hat{\Psi}=\hat{\Psi}^++\hat{\Psi}^-.
\end{equation}
The downstream part is not strongly affected by the black hole creation, hence the particle creation in this part is negligible $\beta_\omega^+=0$. This assumes that the mixing between upstream and downstream parts is small -- otherwise we would have to include a grey-body factor, cf.~\cite{Hawking74}. So Hawking radiation is created by the upstream part only. It can further be decomposed into modes $\Psi_{\omega,\text{super}}^-$ inside the supersonic and $\Psi_{\omega,\text{sub}}^-$ inside the subsonic region.
The decomposition of the field operators is finally
\begin{gather}
 \hat{\Psi}^-=\hat{\Psi}_\text{super}^-+\hat{\Psi}_\text{sub}^-,\\
\hat{\Psi}_\text{super}^-= \int_0^\infty d\omega \kla{\hat{a}_{\omega,\text{super}}^- \Psi_{\omega,\text{super}}^-e^{-i\omega t}+H.c.},\\
\hat{\Psi}_\text{sub}^-= \int_0^\infty d\omega \kla{\hat{a}_{\omega,\text{sub}}^{-\dagger} \Psi_{\omega,\text{sub}}^{-*}e^{-i\omega t}+H.c.}.
\end{gather}
We have to find the regular in field after the horizon formation $\Psi^-_{\omega,\text{in}}$, decompose it in terms of the out modes
\begin{equation}
 \Psi^-_{\omega,\text{in}}\kla{\theta}=\alpha_\omega\Psi^-_{\omega,\text{sub}}\kla{\theta}+\beta_\omega\Psi^-_{\omega,\text{super}}\kla{\theta},
\end{equation}
and read off the Bogoliubov coefficients $\alpha_\omega$ and $\beta_\omega$. So in summary, we calculated the time evolution of global upstream waves before the black hole creation into upstream waves after the black hole creation on both sides of the horizon. In the lab frame the resultant waves travel away from the horizon on both sides.

The interesting Bogoliubov coefficient $\beta_\omega$ is found to be
\begin{equation}
\label{Bogoliubov coeffcients}
 |\beta_\omega|^2=\frac{1}{\exp\kla{\frac{\hbar\omega}{k_\text{B}T_\text{H}}}-1}
\end{equation}
with the Hawking temperature $T_\text{H}$ (see Eq. \eqref{Hawking temperature}). We summarize these findings by stating that Hawking radiation emitted after the creation of a black hole has a thermal spectrum
\begin{equation}
 \bra{\text{in}} \hat N_\omega^\text{out} \ket{\text{in}}=\sum_{\omega}\frac{1}{\exp\kla{\frac{\hbar\omega}{k_\text{B}T_\text{H}}}-1}.
\end{equation}

\subsection{Scattering of Pulses}
\label{Review Scattering}
The derivation of Hawking radiation presented in the last Section \ref{sec.review.mode} suggests calculating the Bogoliubov coefficients from the time evolution of classical pulses scattering at the black hole horizon. It is numerically advantageous to calculate the scattering process backwards in time. In the following we will focus on the situation of a sublinar dispersion relation in a continuum system as studied in \cite{Unruh99}.

The calculation starts from a pulse with negative frequencies and small positive wavenumbers at late times. This pulse is moving upstream and is leftmoving in the comoving as well as in the lab frame. We then calculate its history in time: It approaches the horizon and is reflected by it. The reflected pulse mainly consists of two early-time upstream pulses, one positive and one negative frequency pulse, at high absolute wavenumbers. The positive/negative wavenumber pulse has negative/positive frequencies in the comoving frame, i.e., both are upstream pulses. But they have such a small group velocity - remember the sublinear dispersion relation (see \ref{dispersion}) - that they are dragged along with the ion background. So they are rightmoving in the lab frame, while being leftmoving in the comoving frame. These pulses can be determined due to frequency conservation for a static system in the lab frame. Note that the sign of the frequency in the comoving frame determines the sign of the Klein-Gordon norm (see Eq. \eqref{Klein-Gordon inner product}) and thus the notion of particles versus anti-particles. Therefore, the relation between the late-time negative frequency and early-time positive frequency pulse is important for the particle producing Hawking effect.

\begin{figure*}[tbc]
\begin{center}
\includegraphics[height=45mm,angle=0]{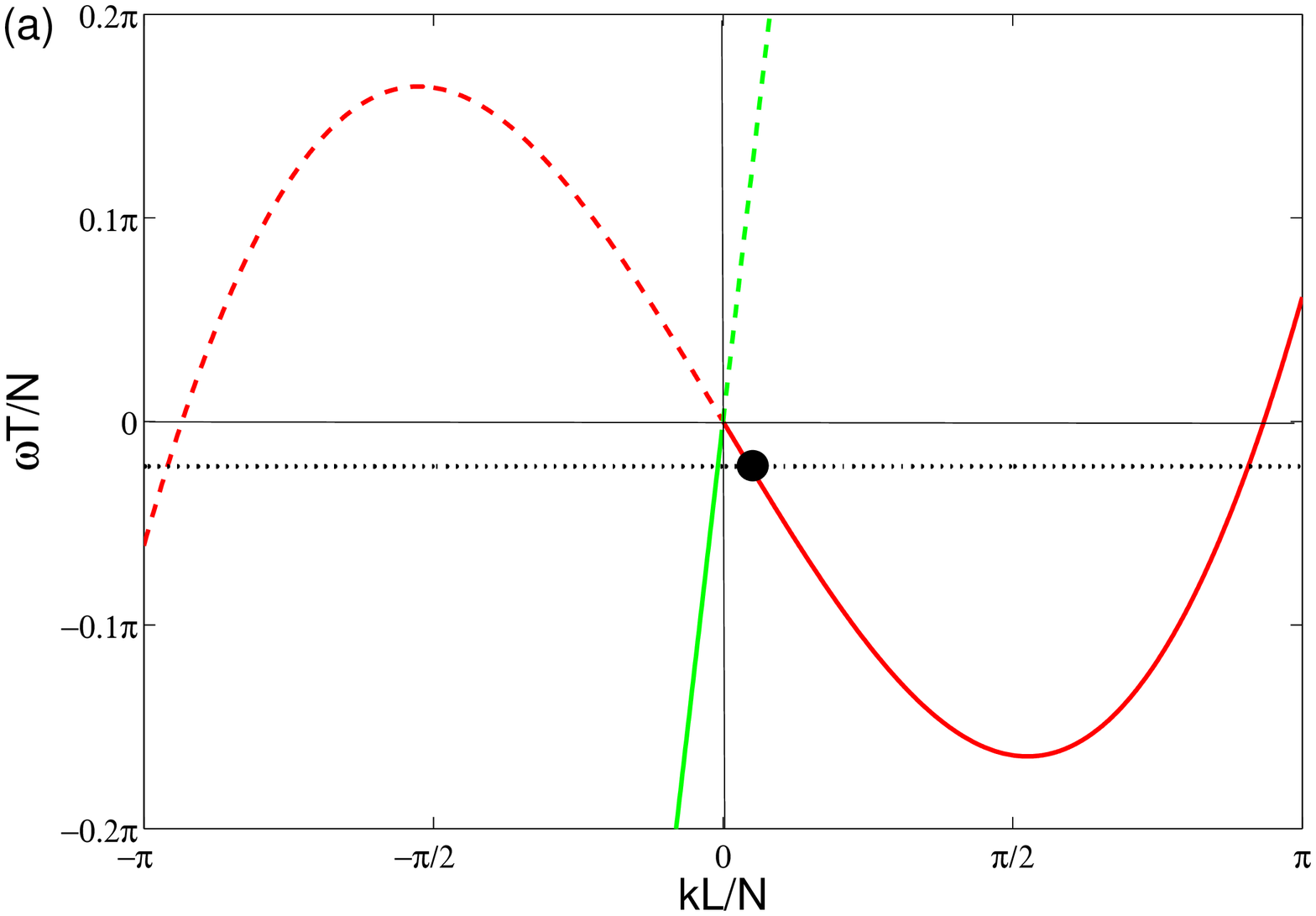}
\includegraphics[height=45mm,angle=0]{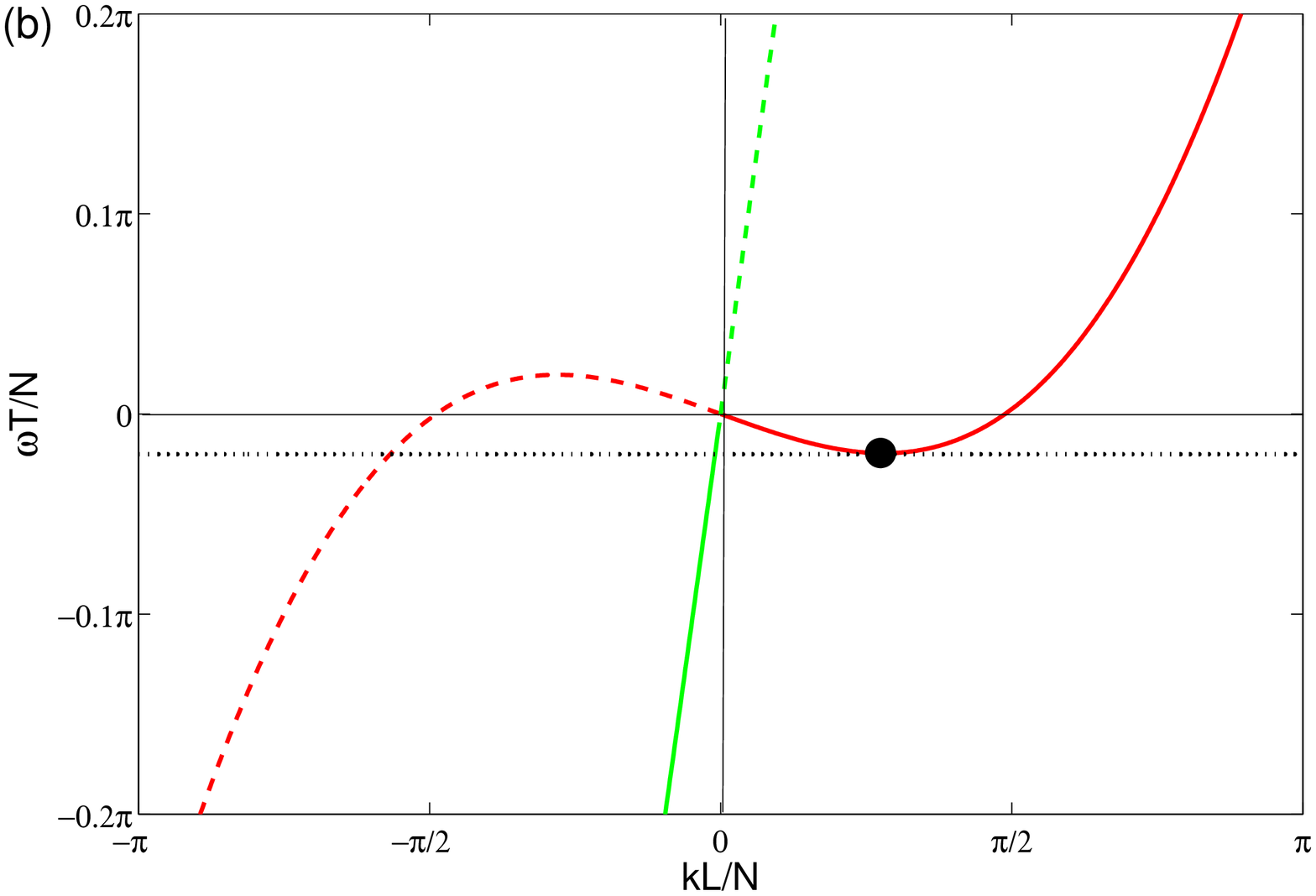}
\includegraphics[height=45mm,angle=0]{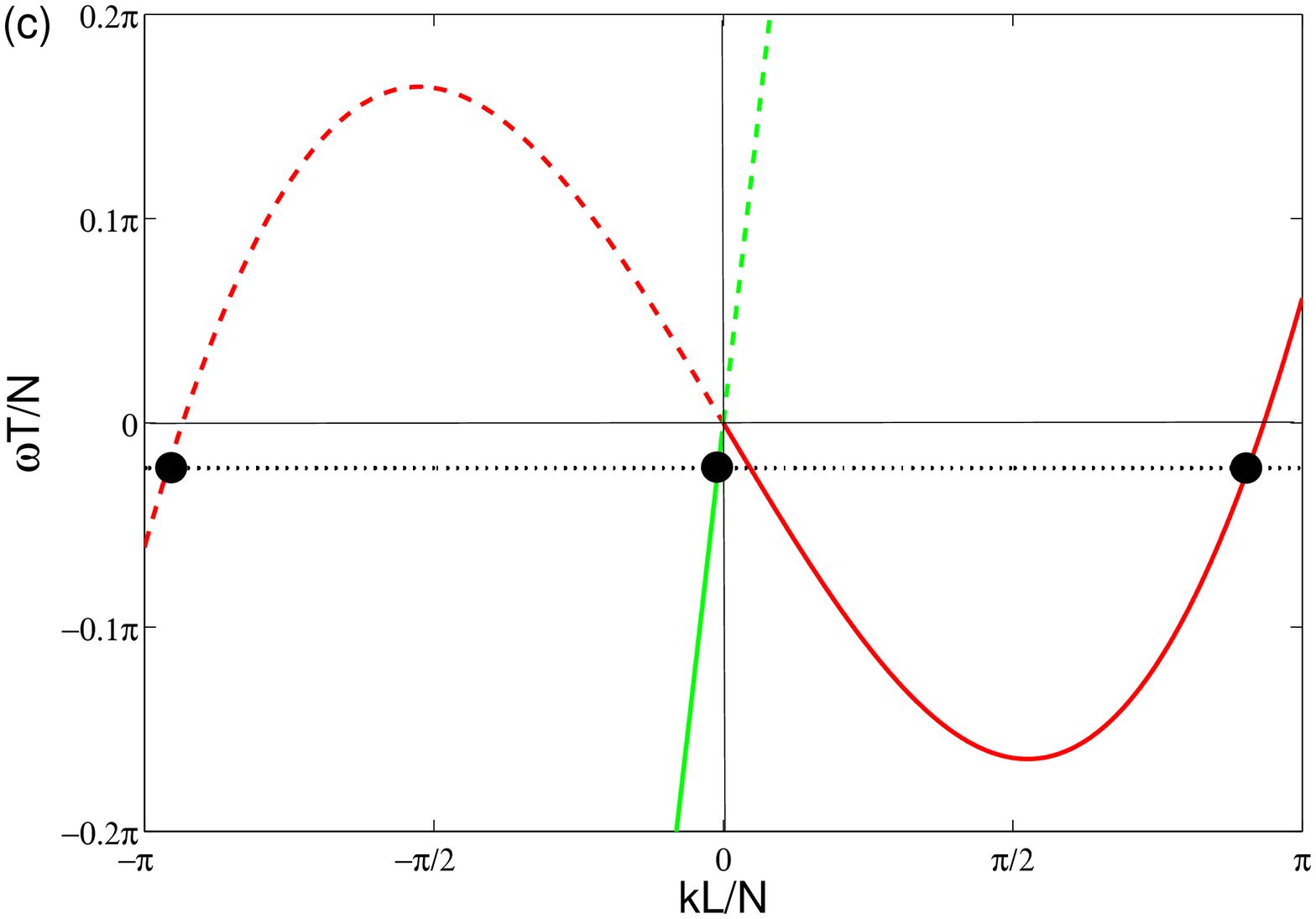}
\includegraphics[height=45mm,angle=0]{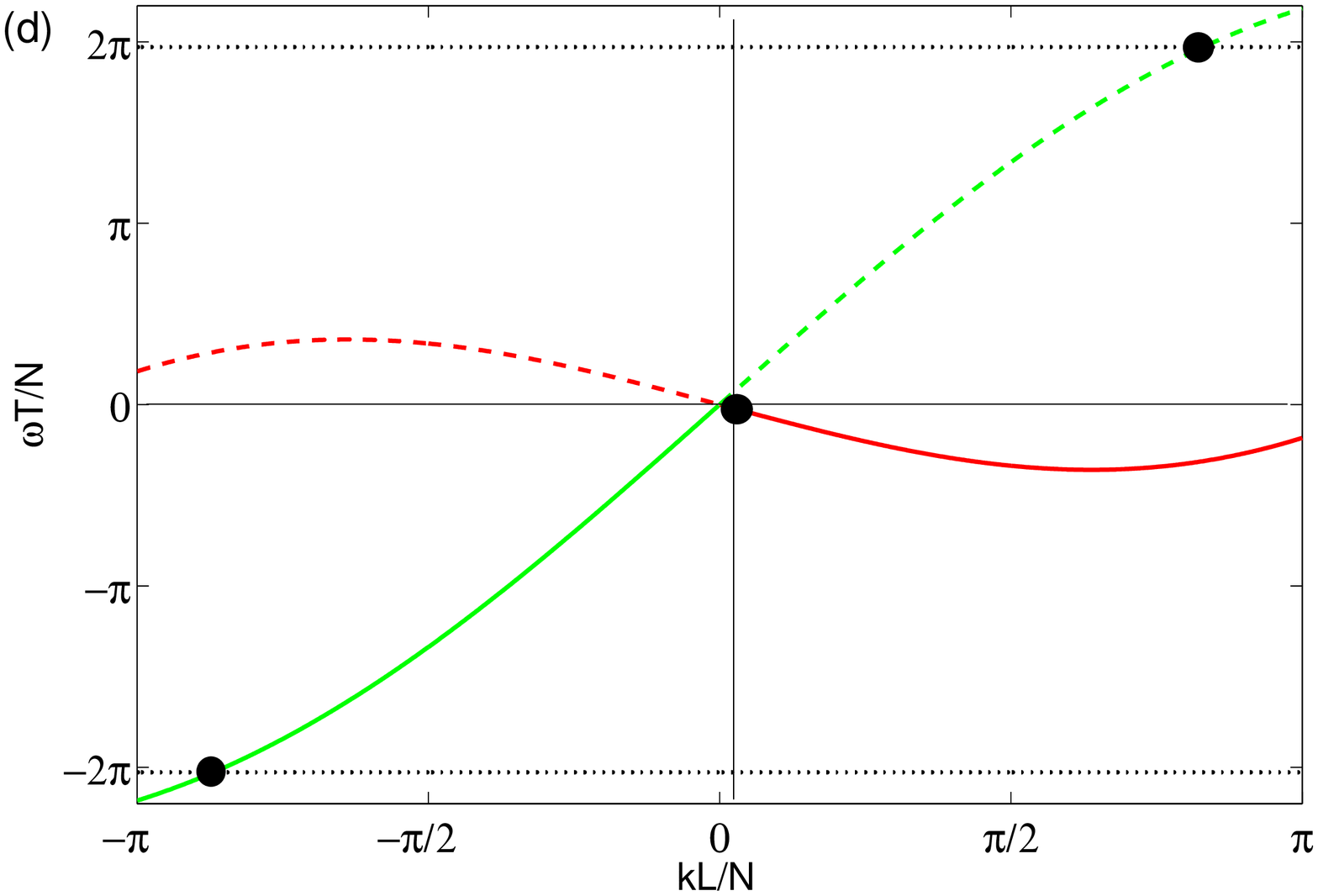}
\caption{Schematical depiction of the dispersion relation for the scattering process of a final negative frequency pulse from the black hole horizon. Negative/positive frequency modes are depicted with solid/dashed lines. Upstream/downstream modes are depicted in red/green. The situation at late times is presented in (a), at intermediate times, when the pulse hits the horizon, in (b), and at early times in (c). Part (d) shows the situation outside of the black hole for small ion velocities, for which we observe the effect analogous to Bloch oscillations (see text).}
\label{dispersion_schematic}
\end{center}
\end{figure*}

Now, we will explain this mode conversion with the help of this frequency conservation and the dependence of the dispersion relation on the local ion velocity. The frequency in the lab frame is
\begin{equation}
\omega_\text{lab}=vk\frac{L}{2\pi}\pm D(k), 
\end{equation}
i.e., it is the Doppler shifted dispersion in the comoving frame $\omega_\text{com}=\pm D(k)$. In Fig. \ref{dispersion_schematic} we depict the pulses at different times during the simulation (a-c) on the dispersion relation in the stationary lab frame together with the initial pulse frequency $\omega_0$. For comparison the interested reader can compare with Fig. \ref{scattering} depicting the pulses in real space. We are starting our discussion at late times (see Fig. \ref{dispersion_schematic}(a)) with a single pulse, upstream and with negative frequency. Its negative group velocity in the lab frame means that it is leftmoving in the lab frame. When the pulse is approaching the black hole horizon backwards in time the ion velocity at the pulse increases. In comparing Fig. \ref{dispersion_schematic}(a) and Fig. \ref{dispersion_schematic}(b) one can observe the blueshifting of the pulse. When the pulse reaches the horizon, it is approximately at the minimum of the dispersion relation (see Fig. \ref{dispersion_schematic}(b)). Now the mode conversion occurs, which is restricted by the frequency conservation in the stationary lab frame. Apart from the late-time pulse, three pulses are in agreement with frequency conservation. One pulse is moving downstream at low negative wavenumbers with negative frequencies. Thus, in the lab frame it is a fast rightmoving pulse. The two other solutions are at high absolute wavenumbers and are upstream, but slowly rightmoving in the lab frame. The positive/negative frequency contribution of the upstream pulse is located at positive/negative wavenumbers. Backwards in time all three pulses travel leftwards away from the horizon (see Fig \ref{dispersion_schematic}(c)). Note that in \cite{Unruh99} the downstream pulse is not observed in the actual dynamics. 

From a comparison between the Klein-Gordon norms of the late-time negative frequency pulse and the early-time positive frequency pulse the Bogoliubov coefficients $|\beta_\omega|^2$ can be extracted. In the literature a different approach is chosen, with the assumption that Hawking radiation is thermal, i.e. that the Bogoliubov coefficient is given by Eq. \eqref{Bogoliubov coeffcients}, the Klein-Gordon norm \ref{Klein-Gordon inner product} of the early-time pulse is calculated from the late-time pulse. The Klein-Gordon norm of the early-time positive frequency pulse is
\begin{equation}
 \mathcal{N}^{+}=\int \mathcal{N}^{+}_k(t<0) dk.
\end{equation}
It is compared with the prediction from the late-time pulse
\begin{equation}
 \mathcal{N}^{0}=\int \frac{\mathcal{N}^{0}_k(t=0)}{\exp\kla{\frac{\hbar\omega}{kT_\text{H}}}-1}dk. 
\end{equation}
Either the norms are compared in total or frequency-wise with the help of the expression $dk=\frac{dk}{d\omega}d\omega$, i.e. by division with the pulse velocities in the lab frame. One can compare the early-time positive frequency pulse
\begin{equation}
\label{spectral1}
 \widetilde{\mathcal{N}}_k^+=\frac{\mathcal{N}_k(t<0)}{v_\text{min}-c_k}
\end{equation}
with the prediction from the late-time negative frequency pulse
\begin{equation}
 \widetilde{\mathcal{N}}_k^0=\frac{\mathcal{N}_k(t=0)}{v_\text{min}-c_k}\frac{1}{\exp\kla{\frac{\hbar\omega_k}{k_\text{B}T}}-1},
\end{equation}
and the prediction from the early-time negative frequency pulse
\begin{equation}
\label{spectral3}
 \widetilde{\mathcal{N}}_k^-=\frac{\mathcal{N}_k(t<0)}{v_\text{min}-c_k}\exp\kla{-\frac{\hbar\omega_k}{k_\text{B}T}}.
\end{equation}
In \cite{Unruh99} the thermal hypothesis is confirmed using both the integrated and the modewise comparison. 

In \cite{Jacobsen99} a discretized hydrodynamic system is treated with the same method. On a lattice the dispersion relation is naturally sublinear. Since the lattice is moving, the frequency in the lab frame is not conserved anymore. It is shown numerically exact and with analytical approximations (WKB theory) that the mechanism of mode conversion described in \cite{Unruh99} persists in this scenario. Even the Bogoliubov coefficients extracted from the comparison of early-time and late-time modes agree with the predictions for a continuum system with strictly linear dispersion relation (see Eq. \eqref{Bogoliubov coeffcients}). The existence of a downstream pulse is not reported in \cite{Jacobsen99}, too.

For a discrete and finite system with $N$ particles on a system of size $L$ a finite number of wavenumbers exists
\begin{equation}
\label{discrete k}
k\in\left\{-\frac{(N'-2)\pi}{L},-\frac{(N'-4)\pi}{L},\dots,\frac{N'\pi}{L}\right\} 
\end{equation}
with the renormalized ion number 
\begin{equation}
\label{renormalized ion number}
N'=N \frac{(2\pi)/T}{v}.
\end{equation}
$N'$ appears because the local ion spacing is not $L/N$, but $a(\theta)=L/N'=L/N\cdot v(\theta)T/(2\pi)$ in the inhomogeneous system. This Brillouin zone is already visualized in Fig. \ref{dispersion_schematic}. In this system a symmetry under combined translations in space and time survives, 
\begin{gather}
\label{higher symmetry}
i\rightarrow i+1\\
t\rightarrow t+T/N
\end{gather}
(see Eq. \eqref{definition g}). In a homogeneous part of the system this symmetry implies that the state of the system is invariant under the combined transformation
\begin{gather}
 k\rightarrow k+\frac{2\pi}{L} N' n,\\
 \omega\rightarrow \omega+\frac{2\pi}{T}Nn
\end{gather}
for any integer $n$. 

We use this transformation to explain an effect described in \cite{Jacobsen99} analogous to Bloch oscillations. The two high wavenumber solutions of the frequency conservation condition can lie outside the Brillouin zone. Then we find the solutions by looking at the frequencies $\omega_0\pm 2\pi N/T$, depicted in Fig. \ref{dispersion_schematic}(d). Thus, the early-time pulses are instead located on the downstream branch of the frequency condition. The positive/negative frequency pulse is located at high positive/negative wavenumbers (see Fig. \ref{dispersion_schematic}(d)). 

\subsection{Correlations}
\label{Review Correlations}
The Hawking effect produces pairs of particles propagating away from the horizon in opposite directions, one outside the black hole (the Hawking particle), and one inside the black hole (its infalling partner). Taking advantage of this, it has been proposed to detect Hawking radiation via correlation measurements (see \cite{Carusotto08}), which reveal the entanglement between the two Hawking partners. In \cite{Carusotto08} the density-density cross-correlation of phonons on both sides of the horizon produced through the Hawking process in a sonic black hole built up with a Bose-Einstein condensate (BEC) is studied \cite{Cirac00}. In a quasi one-dimensional weakly interacting BEC in the hydrodynamical approximation the sound propagation is described by
\begin{equation}\label{L.sound}
 \mathcal{L}=\int d\theta\frac{\mathcal{K}}{2 }\left[\kla{\partial_t\Phi+v\kla{\theta}\partial_{\theta}\Phi}^2-c^2\partial_\theta\Phi^2\right].
\end{equation}
This is manifestly the same as Eq. \eqref{analogconstantc} apart from the actual form of the conformal factor $\mathcal{K}$, which if constant does not affect the dynamics appart from a rescaling of the fields. So the propagation of phonons in a BEC, i.e. phase/density excitations, and the propagation of phonons in a ring of ions, i.e. displacements of ions from their equilibrium position, are described by the same physics, and therefore share similar behaviors. Thus, the analysis performed in \cite{Carusotto08} can be translated into the context of ion rings. 

The general equal-time two-point correlator $\langle \hat\Psi(\theta) \hat\Psi(\theta')\rangle$ in 1+1 dimensional space-times is given by \cite{BD}
\begin{equation}
\label{field.corr.gen}
\langle \hat\Psi(\theta)\hat\Psi(\theta')\rangle=-\lim_{t'\rightarrow t} \frac{\hbar}{4\pi
}\log \left[\left(U(\theta,t)-U(\theta',t')\right)
\left(V(\theta,t)-V(\theta',t')\right)\right].
\end{equation}
This is the stationary correlator of a massless minimally coupled quantum scalar field in 1 spatial dimension. It exhibits a characteristic structure with the logarithmic distance between the two points as the difference between the null coordinates $U$ and $V$ \cite{BD}. The meaning of the null coordinates is the following: the mode solution can always be written in the form $\sim e^{-i\omega U}$ for upstream modes and $\sim e^{-i\omega V}$ for downstream modes. The modes propagating upstream (downstream) move along space-time trajectories that keep the $V$ ($U$) coordinates constant. For example, if the system is at rest and homogeneous, the analog spacetime is the Minkowskii (flat) one, and the Kruskal coordinates reduces to the familiar $U=t+\theta/c$, $V=t-\theta/c$  \cite{note}, that define the standard light- (sound-) cones. If the system is more complicated (i.e. non-homogeneous or moving), the mode propagation is different, and the associated null coordinates will display distortion. 

Eq. \eqref{field.corr.gen} applies to all conformally invariant theories in 1+1 dimensional spacetimes (which are always conformally flat). In the present case, the theory we are dealing with is not conformally invariant for the presence of the conformal factor $\mathcal{K}$ in the action. Nevertheless, assuming that $\mathcal{K}$ varies smoothly over the system, the correlator can be approximated using Eq. \eqref{field.corr.gen} also in the present case by 
\begin{equation}
\label{field.corr.dilaton}
\langle\hat\Phi(\theta)
\hat\Phi(\theta')\rangle=-\lim_{t'\rightarrow t} \frac{\hbar}{4\pi
}\frac{1}{\sqrt{\mathcal{K}(\theta)\mathcal{K}(\theta')}} 
\log \left[\left(U(\theta,t)-U(\theta',t')\right) \left(V(\theta,t)-V(\theta',t')\right)\right],
\end{equation}
where the $1/\sqrt{\mathcal{K}(\theta)\mathcal{K}(\theta')}$ term follows from a rescaling of the field, and terms containing derivatives of $\mathcal{K}$ are neglected.

Let us move to the evaluation of this correlator in the presence of Hawking radiation. Since the modes responsible for the emission of Hawking radiation are the upstream modes $\hat\Psi^-$, we will focus on the upstream sector of Eq. \eqref{field.corr.gen} only. The equal-time two-point correlator in the presence of a black hole horizon is given by \cite{BD}
\begin{equation}\label{field.corr.BH}
\langle \hat\Phi^-(\theta)\hat\Phi^-(\theta')\rangle=-\lim_{t'\rightarrow t} \frac{\hbar}{4\pi
}\frac{1}{\sqrt{\mathcal{K}(\theta)\mathcal{K}(\theta')}} 
\log\left(U(\theta,t)-U(\theta',t')\right).
\end{equation}
The downstream part of the correlator remains unaffected even in the presence of a horizon, i.e. $V$ still reads $t-\theta/c$. The only modes which get distorted by the presence of the horizon are the upstream ones which become far away from the horizon \cite{Unruh81, backreaction, Balbinot2006} 
\begin{equation}\label{U}
U(\theta, t)_{in/out}=\pm e^{-\kappa
(t+\frac{\theta}{c(\theta)-v(\theta)})}
\end{equation}
for the interior (+) and the exterior (-) region of the black hole. They suffer the typical exponential distortion of the upstream modes due to the presence of a black hole horizon \cite{BD,Unruh81} (discarding transients). The exponential distortion follows from the wave equation for a linearized velocity profile at the horizon. $\kappa$ is the surface gravity on the sonic horizon which is proportional to the Hawking temperature (see Eq. \eqref{Hawking temperature}). It is worth emphasizing that the form of the modes (\ref{U}) is universal for any black hole horizon formation, independent of the details of its formation. This is the origin for the universal behavior of Hawking radiation.

The momentum-momentum correlator $\langle \delta\hat p^-(\theta)\delta \hat p^-(\theta')\rangle$ can be obtained from Eq. \eqref{field.corr.BH}. The conformal factor $\mathcal{K}$ must be replaced with the conformal factor from the Lagrangian of ions on a ring (see Eq. \eqref{analogconstantc}). With the relation
\begin{equation}
\label{momentum_angle}
\delta \hat p^-=\frac{L}{2\pi}m(\partial_t+v\partial_\theta)\hat\Phi^-
\end{equation}
one gets
\begin{multline}
\label{momentum.corr} \langle \delta \hat p^-(\theta) \delta\hat
p^-(\theta')\rangle =\frac{\hbar m}{16\pi}
\frac{1}{\klab{c\kla{\theta}-v\kla{\theta}}\klab{c\kla{\theta'}-v\kla{\theta'}}}\cdot\\
\sqrt{\frac{c(\theta)c(\theta')}{n(\theta)n(\theta')}}
\frac{\kappa^2}{\cosh^2\left[ \frac{\kappa}{2}\left(
\frac{\theta}{c\kla{\theta}-v\kla{\theta}} -
\frac{\theta'}{c\kla{\theta'}-v\kla{\theta'}}\right)\right]}.
\end{multline}
This correlator has the typical form associated to the Hawking effect. The correlations are scaling with the square of the Hawking temperature. For $\theta$ and $\theta'$ on opposite sides of the sonic horizon, $c\kla{\theta}-v\kla{\theta}$ and $c\kla{\theta'}-v\kla{\theta'}$ have opposite sign. Therefore, the
momentum-momentum cross-correlations are negative. They exhibit a peculiar peak along a straight line for
$\theta=\frac{c\kla{\theta}-v\kla{\theta}}{c\kla{\theta'}-v\kla{\theta'}}\theta'\propto 
\theta'$. These cross-correlations correspond to the two entangled Hawking particles propagating in opposite directions as they move apart from the horizon.

\section{Simulations for Ion Rings}
\label{simulations}
\begin{figure*}[t!!!]
\begin{center}
\includegraphics[width=75mm,height=56mm,angle=0]{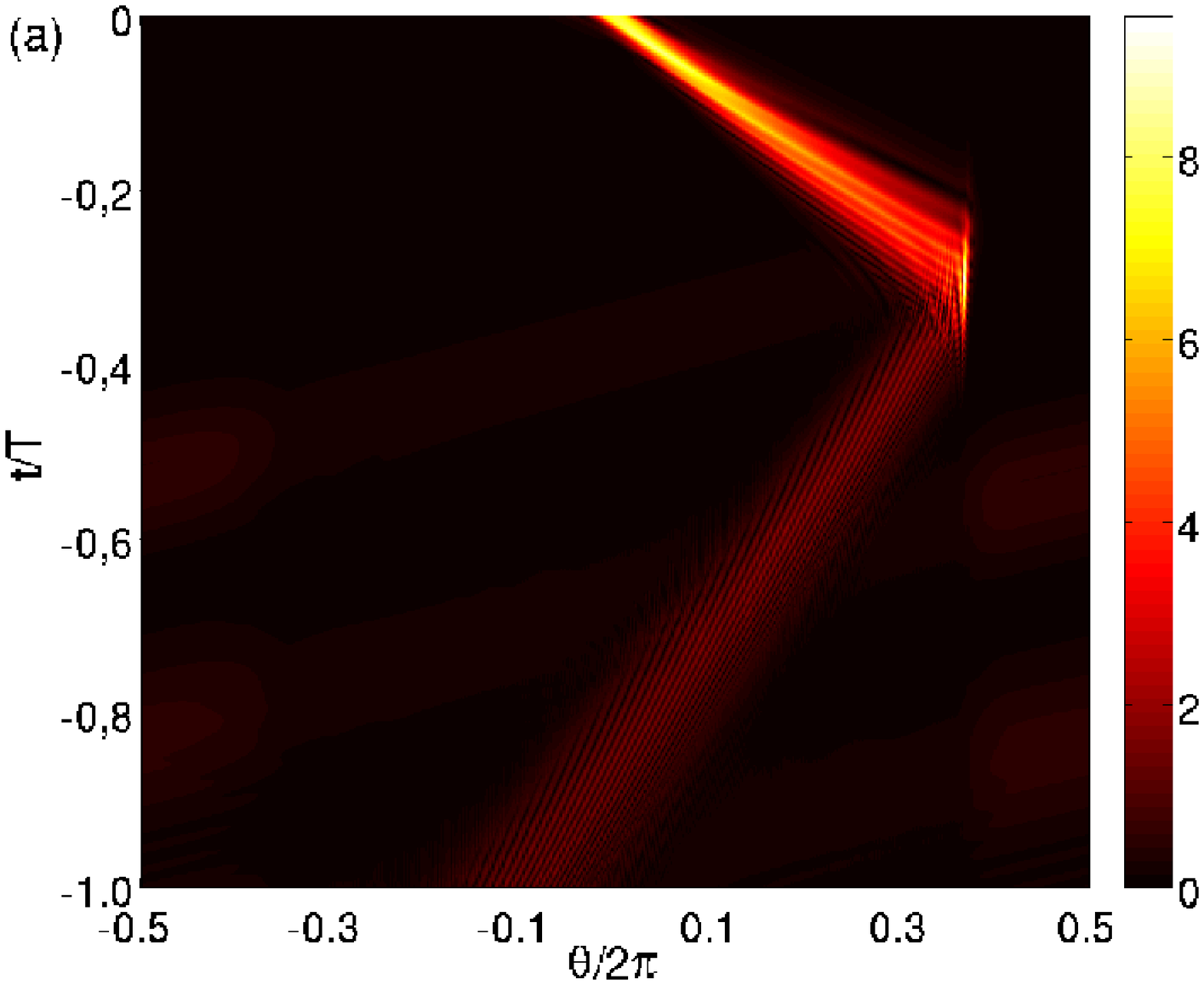}
\includegraphics[width=75mm,height=56mm,angle=0]{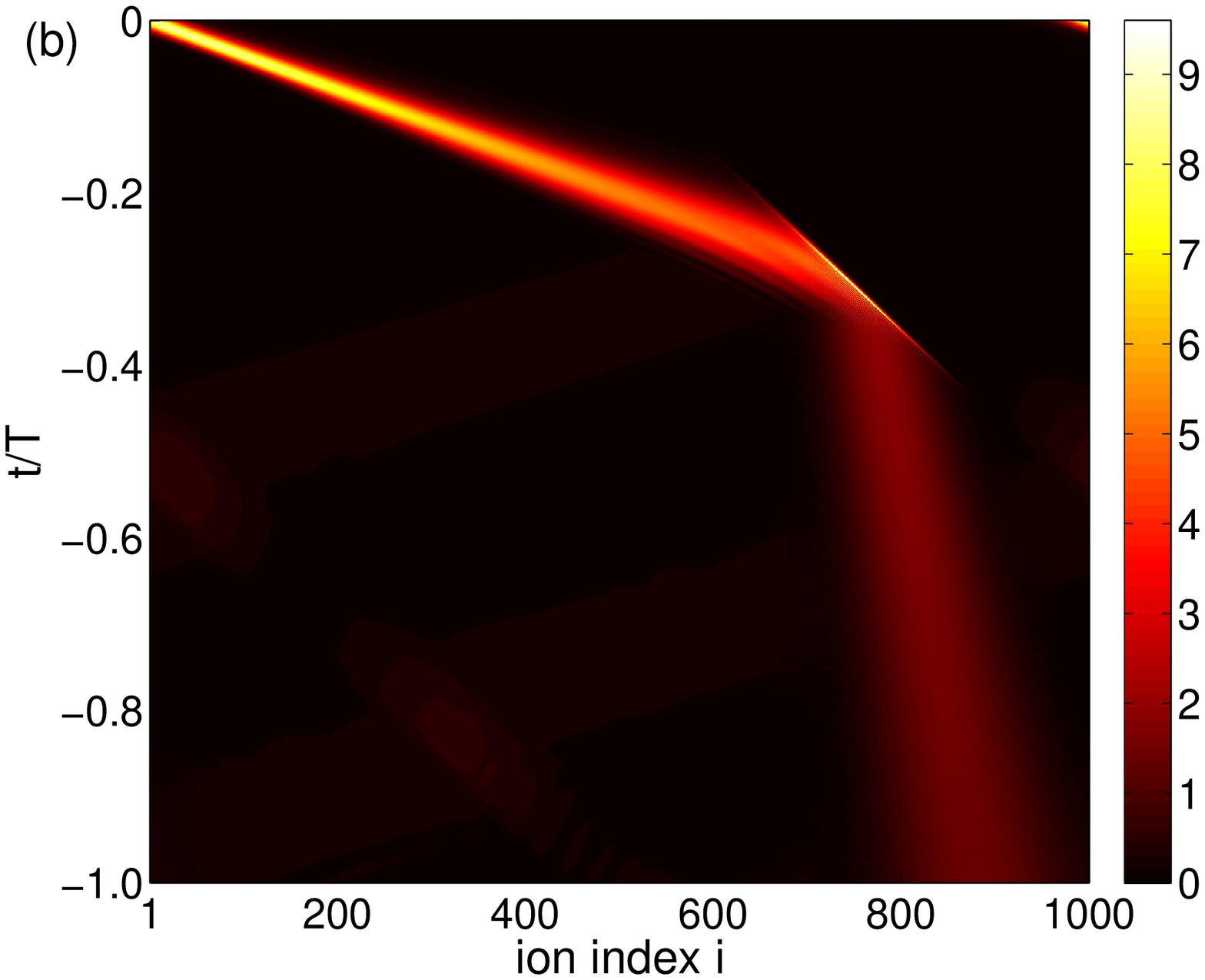}
\includegraphics[width=75mm,height=56mm,angle=0]{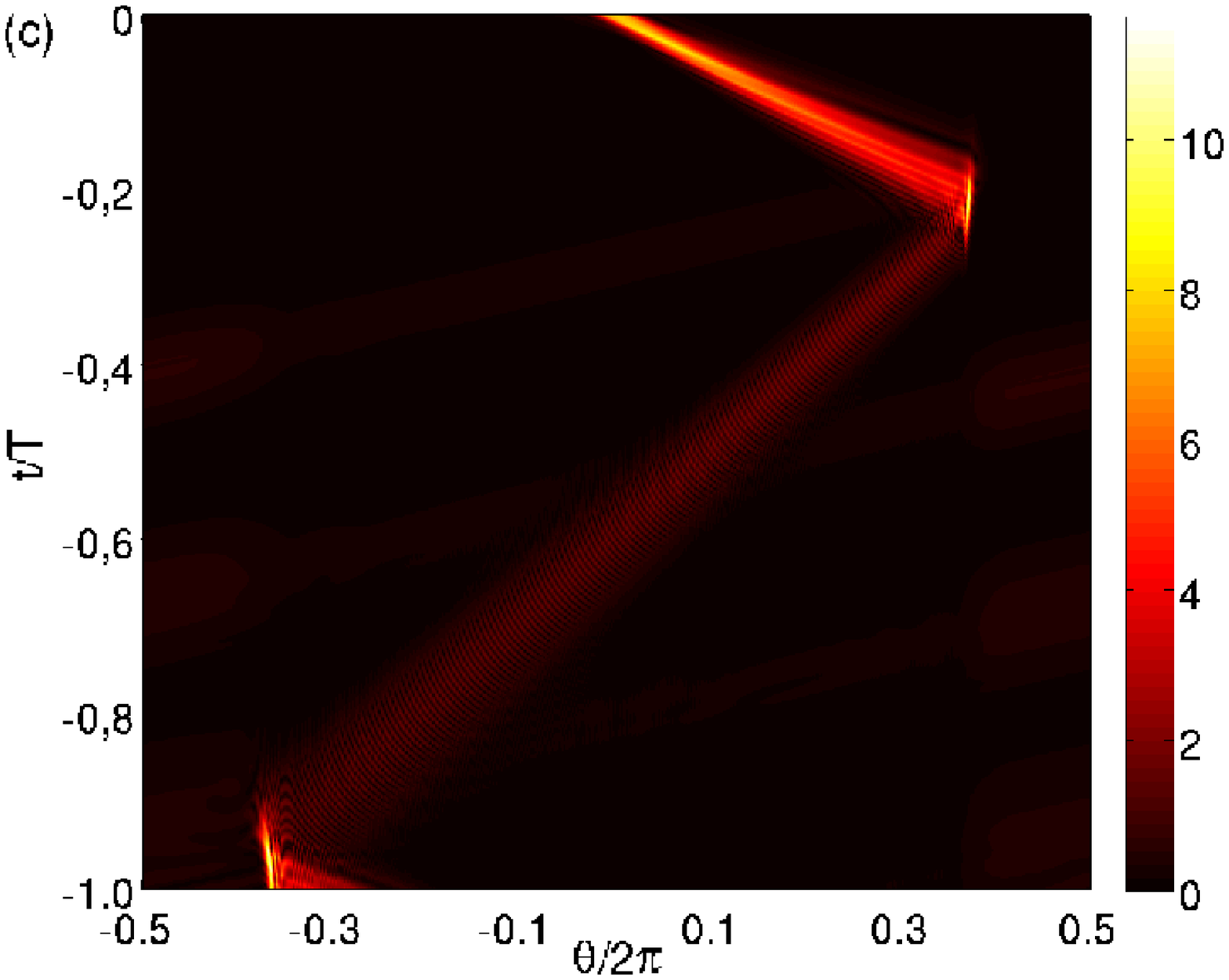}
\includegraphics[width=75mm,height=56mm,angle=0]{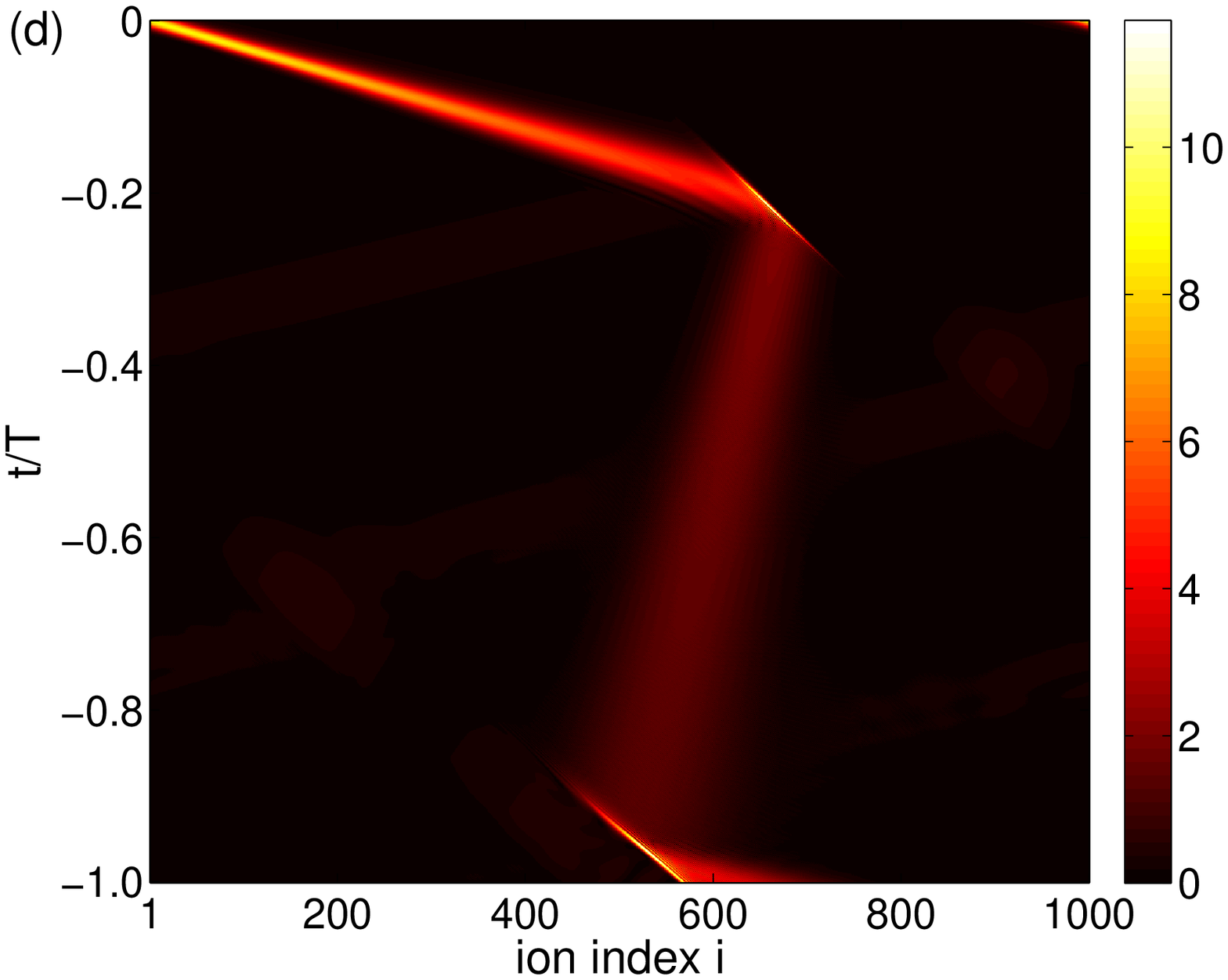}
\caption{Ion displacement $|\delta\theta_i(t)/(2\pi)|$ during propagation backwards in time starting from the final wavefunction in Eq. \eqref{final} with $s=5$. One late-time pulse and three (two of which overlapp) early-time pulses are present (see text). Two early-time pulses have large absolute momenta (blue dashed lines), one early-time pulse has small momentum (green straight lines). We use $\sigma v_\text{min}T=0.375$ and $N=1000$. For (a) and (b) $e^2/4\pi\epsilon_0=\frac{1.2591}{2N} \frac{mL^3}{T^2}$, for (c) and (d) $e^2/4\pi\epsilon_0=\frac{2.0004}{2N} \frac{mL^3}{T^2}$ (see \ref{section velocity profile}). (a) and (c) use lab frame angles $\theta$, (b) and (d) use ion indices $i$ (see text for description of scattering process.}
\label{scattering}
\end{center}
\end{figure*}
We are now returning to the discussion of the discrete ion chain. In this section we are presenting the results of our simulations and are comparing them with the predictions and expectations from Sec. \ref{Review}. We are pursuing two routes of simulations: First we simulate the scattering of pulses on the black hole horizon in Sec. \ref{sec.scattering}. From the result we can deduce the Bogoliubov coefficients and theoretically confirm the thermal hypothesis (see Sec. \ref{Review Scattering}). A second series of simulations presented in Sec. \ref{sec.correlations} is analyzing the emergence of correlations between the inside and the outside of a black hole after its creation (see Sec. \ref{Review Correlations}). These correlations demonstrate the pair creation mechanism of Hawking radiation and are closely related to the emergence of entanglement between the inside and the outside of a black hole. In contrast to the scattering analysis, these simulations also act as direct proposal for an experiment as further discussed in Sec. \ref{Experimental Parameters}.

\subsection{Scattering of Pulses}
\label{sec.scattering}
\begin{figure}[tbc]
\begin{center}
\includegraphics[width=75mm,angle=0]{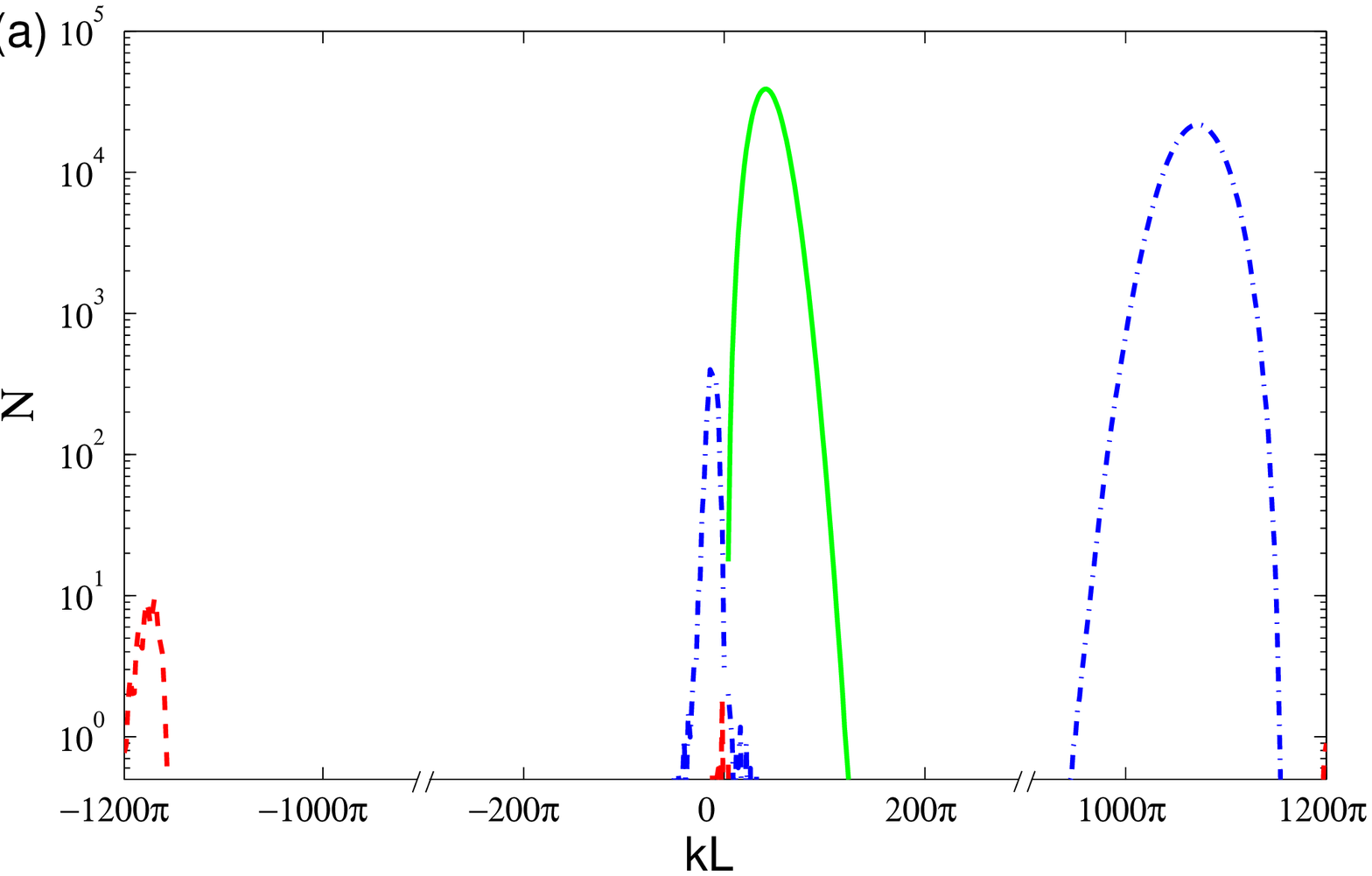}
\includegraphics[width=75mm,angle=0]{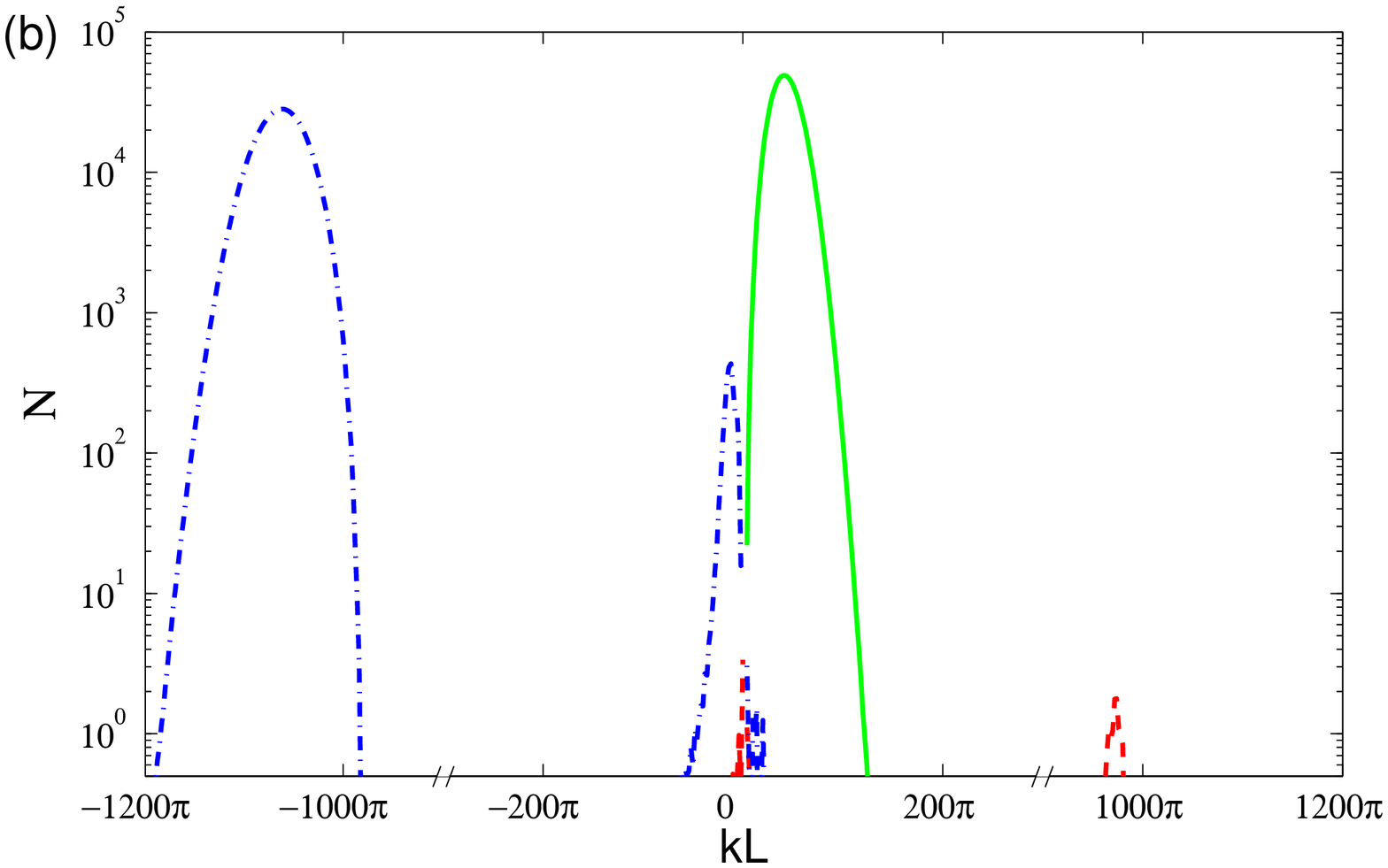}
\caption{Klein-Gordon norm $\mathcal{N}_k$ of the phonon pulses for propagation backwards in time starting from the final wavefunction in Eq. \eqref{final} with $s=5$. With a comparison of these pulses we test the thermal hypothesis. We use $\sigma v_\text{min}T=2\pi\cdot 0.375$, and $N=1000$ (see \ref{section velocity profile}). The final negative frequency pulses are depicted in green, the initial negative frequency pulses in blue (dash-dotted line), and the initial positive frequency pulses in red (dashed line). We depict the discrete norm distributions/dispersion relation as a continuous curve here. (a) Final ($t=0$) and initial ($t=-0.67T$) norm distribution of $\delta\theta_k(t)$; $e^2/4\pi\epsilon_0=1.2591/(2N)\cdot mL^3T^{-2}$. (b) Final ($t=0$) and initial ($t=-0.51T$) norm distribution of $\delta\theta_k(t)$ showing Bloch oscillation; $e^2/4\pi\epsilon_0=2.0004/(2N)\cdot mL^3T^{-2}$ \cite{Horstmann10}.}
\label{fourierspace}
\end{center}
\end{figure}
Our numerical results for the propagation of a final pulse backwards in time in the discrete system of phonons on an ion ring are presented in the following. We first introduce the quantities necessary for this analysis. If the phononic excitations are localized in the flat subsonic region with constant ion velocity $v_\text{min}$, the excitations $\delta\theta_i(t)$ and $\delta\dot\theta_i(t)$ can be expressed as modes $\delta\theta_k(t)$ and $\delta\dot\theta_k(t)$ with wavenumber $k$, where the dot represents the time derivative in the comoving frame. Due to the finite system size only discrete wavenumbers appear (see Eqs. \eqref{discrete k} and \eqref{renormalized ion number}) and thus the dispersion relation is also discrete. The positive and negative frequency part of these excitations are defined by
\begin{gather}
 \delta\theta_k^\pm(t)=\frac{1}{2}\kla{\delta\theta_k(t)\pm i \delta\dot\theta_k(t)/\omega_k},\\
 \delta\dot\theta_k^\pm(t)=\frac{1}{2}\kla{\delta\dot\theta_k(t) \mp i \omega_k \delta\theta_k(t)}.
\end{gather}
These relations follow from the spatial behavior $\delta\theta\kla{\theta}\sim\exp\kla{ik\frac{\theta}{2\pi}L}$ in regions of constant ion velocity $v\kla{\theta}=v$ \cite{Unruh99}. The analysis of the particle production requires us to use the Klein-Gordon norm for these modes. In our special case the Klein-Gordon norm defined in Eq. \eqref{Klein-Gordon inner product} becomes up to a constant
\begin{equation}
 \mathcal{N}=\sum_{k,\pm} \mathcal{N}_k
\end{equation}
with
\begin{equation}
\mathcal{N}_k^\pm=\delta\dot\theta_k^{\pm *}\delta\theta^\pm_k-\delta\theta_k^{\pm *}\delta\dot\theta^\pm_k.
\end{equation}
We are now describing the numerical calculation of the Bogoliubov coefficients with the method presented in Sec. \ref{Review Scattering}. In summary, we are calculating the history of a negative frequency pulse on the upstream branch of the dispersion relation that travels away from the horizon. Back in time it scatters off the horizon and originates from several pulses. The early-time positive frequency pulse and the late-time negative frequency pulse are related through the Bogoliubov coefficients.

We start from the final pulses
\begin{equation}
\label{final}
 \delta\theta^s_k\kla{0}=k \cdot e^{-\kla{\frac{k-2\pi s}{40\pi}}^2}, \hspace{0.3 cm} s=1,\dots,20,
\end{equation}
centered at different wavenumbers $k$ to test different frequency ranges. We calculate its history with Newton's equations of motion given in \ref{equations of motion} (see Eq. \eqref{Newton}) by using an iterative differential equation solver. From frequency conservation in the lab frame we expect three pulses on the upstream branch of the dispersion relation \cite{Unruh99,Jacobsen99} (see Sec. \ref{Review Scattering}).

\begin{figure}[tbc]
\begin{center}
\includegraphics[width=75mm,height=57mm,angle=0]{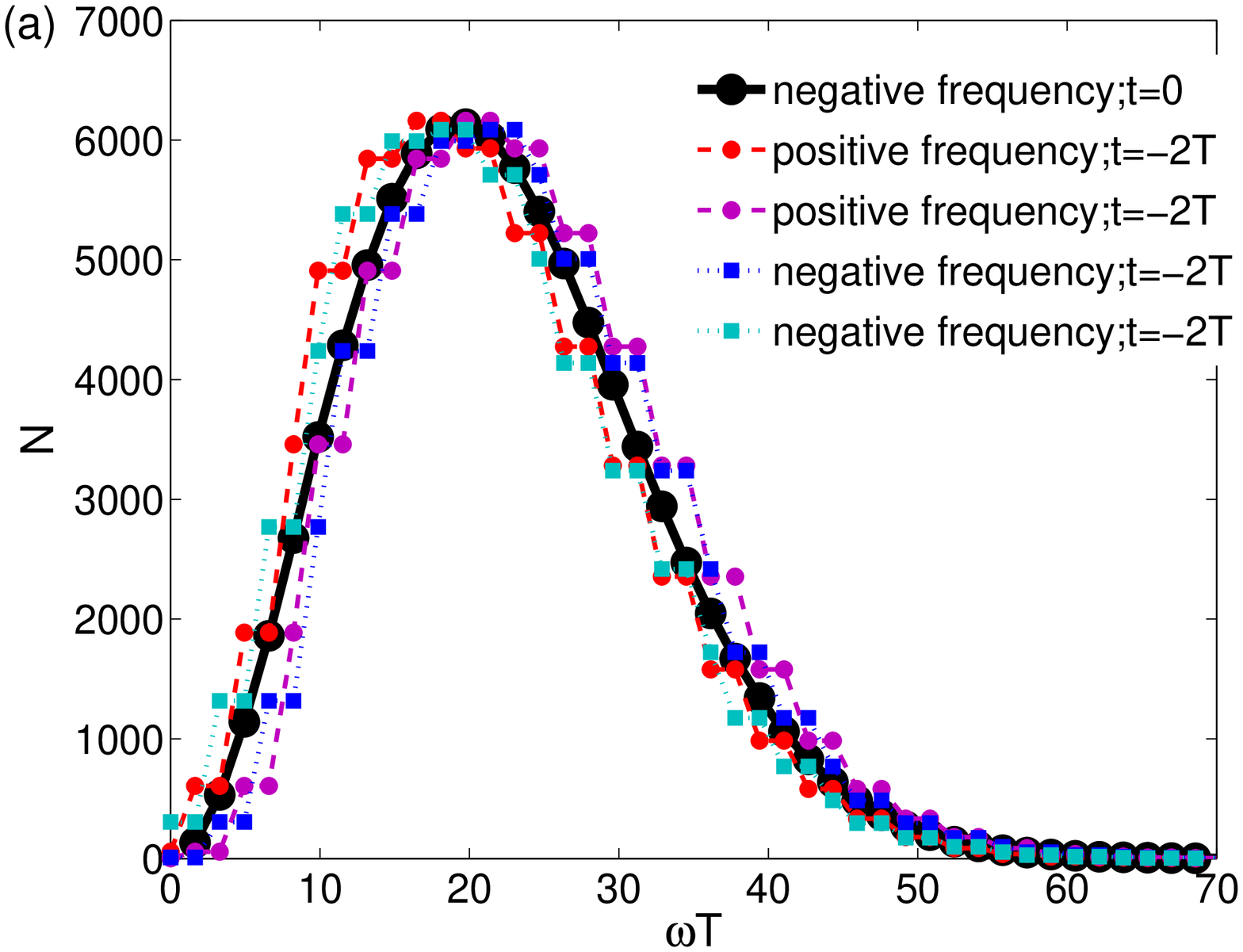}
\includegraphics[width=75mm,height=57mm,angle=0]{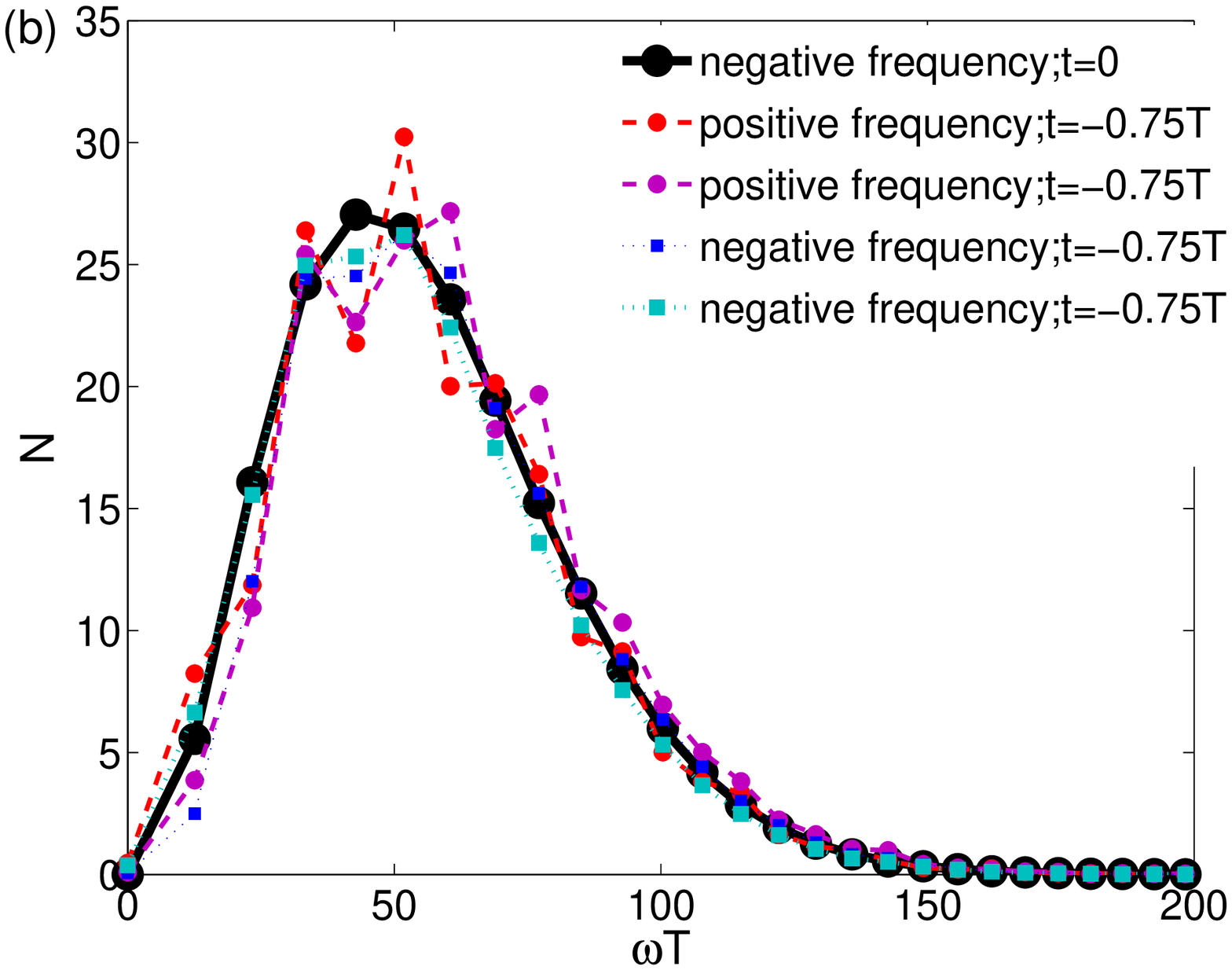}
\caption{Comparison of spectrally resolved Klein-Gordon norms $\widetilde{\mathcal{N}}_k^0$ (late-time negative frequency pulse) at $t=0$ with $\widetilde{\mathcal{N}}_k^+$ (early-time positive frequency pulse) and $\widetilde{\mathcal{N}}_k^-$ (early-time negative frequency pulse) (see Sec. \ref{Review Scattering}) as a function of the lab frame frequencies at $t=-2T$ after propagation backwards in time starting from the final wavefunction in Eq. \eqref{final} with $s=5$. We use $\sigma v_\text{min}T=0.375$, $N=1000$. (a) Nearest-neighbor interactions at $t=-2T$ with $e^2/4\pi\epsilon_0=\frac{1}{2N}\frac{mL^3}{T^2}$; (b) Full Coulomb interactions at $t=-0.75T$ with $e^2/4\pi\epsilon_0=\frac{1}{2N}\frac{mL^3}{T^2}$ (see \ref{section velocity profile}).}
\label{modewise comparison}
\end{center}
\end{figure}

Before coming to the simulation results, we will discuss the parameter regime used in this section. The ratio between the typical pulse frequency $\omega$ in the comoving frame and the Hawking temperature is of order $\hbar\omega/k_BT_H\sim 10$ (see also Fig. \ref{temperature}). In \cite{Macher09} the deviations from thermality have been examined based on the quantity $\omega_\text{max}=\max_{k>0}\kla{D(k)-vkL/2\pi}$. Here $\omega_\text{max}$ takes the following values: $\omega_{max}=290/T$ for nearest-neighbor interactions only (Fig. \ref{modewise comparison}a), $\omega_{max}=610/T$ for full Coulomb interactions (Fig. \ref{modewise comparison}b). The surface gravity for the former case is $\kappa=65/T$, for the latter $\kappa=82/T$ (see Fig. \ref{temperature}), thus the ratios are $\omega_{max}/\kappa=4.5$ and $\omega_{max}/\kappa=7.4$. This regime is identified as the regime of small deviations from thermality in \cite{Macher09}. This prediction applies to systems with nearest-neighbor interactions only. One of the main results of our paper is to consider also the long range Coulomb interactions for which the deviations from a linear dispersion relation at small wavenumbers are significant in a finite system.

Our results agree mainly with those of references \cite{Unruh99,Jacobsen99} as shown in Fig. \ref{fourierspace}. First, we observe quantitative devisions, whose order of magnitude agrees with the uncertainties in calculating the Hawking temperature. Second, we find a previously undescribed downstream pulse at small negative frequencies. But before discussing the detailed analysis of the scattering process, we will repeat the description of the mode conversion from Sec. \ref{Review Scattering}, but this time illustrated in real space with the simulation results from the special case of the ion ring (see Fig. \ref{scattering}). 

We begin with the normal scenario of sufficiently large ion velocities (see Figs. \ref{scattering}(a)-(b) and \ref{fourierspace}(a)). We observe all three early-time pulses which are in agreement with frequency conservation. Two pulses are located on the upstream branch at high absolute wavenumbers (red and blue lines in Fig. \ref{fourierspace}(a)). But at these wavenumbers they have such a small group velocity in the comoving frame that they are moving rightwards in the lab frame (blue dashed lines in Figs. \ref{scattering}(a)-(b)). A third pulse is located on the downstream branch of the dispersion relation at small negative wavenumbers (blue line in Fig. \ref{fourierspace}(a)). Its group velocity is large both in the comoving frame and in the lab frame. This time evolution is depicted in Fig. \ref{scattering}(a) in the lab frame and in Fig. \ref{scattering}(b) in the comoving frame. The two upstream and the one downstream pulse can clearly be identified in the comoving frame. In the lab frame the downstream pulse is quickly moving rightwards, the two upstream pulses are slowly moving rightwards.

A different situation arises for sufficiently small ion velocities (see Figs. \ref{scattering}(c)-(d) and \ref{fourierspace}(b)). In this case, all three early-time pulses are moving downstream. This time evolution is depicted in Fig. \ref{scattering}(c) in the lab frame and in Fig. \ref{scattering}(d) in the comoving frame. In the comoving frame we can now observe the three downstream pulses, one fast pulse which is the one at low wavenumbers (green lines) and the two interesting slow pulses (red dashed lines) which are still located at high absolute wavenumbers. In the lab frame the two pulses at high absolute wavenumbers are travelling faster than in the normal case.

The spectral analysis in Fig. \ref{fourierspace} confirms these explanations. We find the two previously mentioned high wavenumber pulses. For small ion velocities the effect analogous to Bloch oscillations is observed. Then the main incoming pulses have wavenumbers with opposite signs. In addition we observe the weak downstream pulse with small negative wavenumbers that is not described in the literature. We find that its magnitude depends on the strongly on how the fringes of the pulse in outside the flat subsonic region are treated. Therefore we suggest that the presence of this pulse could be a consequence of the finite initial and final excitation probability of ions outside the flat subsonic region.

We further compare the Klein-Gordon norm of the positive frequency early-time pulse with the prediction for thermal radiation (see Sec. \ref{Review Scattering} and \cite{Unruh99,Jacobsen99}). We calculate the Hawking temperature according to Eq. \eqref{temperatureLDA}. If only nearest-neighbor Coulomb interactions are considered, the relative difference between these norms is lower than $\epsilon=0.01$ for $N=1000$ ions. For the long range Coulomb interactions a conservative estimate yields $\epsilon\le 0.2$. The latter result holds for any arbitrary choice for the Hawking temperature at the wavenumbers $k=2\pi/L\dot4\dots 6$. The bound on $\epsilon$ agrees with the differences between analysis based on the group and the phase velocity of the phonons (see Sec. \ref{section continuum limit} and Eq. \eqref{phase group velocity}).

We also perform the analogous spectrally dissolved comparison (see Sec. \ref{Review Scattering} and \cite{Unruh99}), comparing the early-time positive frequency pulse with the prediction based on the thermal hypothesis. Only discrete frequencies appear in the system because only discrete wavenumbers are present. Therefore, it is in general not possible to exactly match the frequencies of the pulses. Especially, we cannot perform a thermal fit. Thus, we look at the two frequencies in the early-time pulses closest to the frequencies in the late-time pulse. We compare the early-time positive frequency pulse $\widetilde{\mathcal{N}}_k^+$ with the predictions for it based on the late-time negative frequency pulse $\widetilde{\mathcal{N}}_k^0$ and based on the early-time negative frequency pulse $\widetilde{\mathcal{N}}_k^-$ (see \eqref{spectral1}-\eqref{spectral3}). The result of this analysis is shown in Fig. \ref{modewise comparison}(a) for nearest-neighbor interactions and in Fig. \ref{modewise comparison}(b) for full Coulomb interactions. For this analysis we choose the Hawking temperature at $k=2\pi/L\cdot 5$ (see Eq. \eqref{temperatureLDA}), but similar results are obtained for the adjacent wavenumbers). This analysis again confirms the thermal hypothesis to the extend possible. The accuracy of our analysis is restricted by the discreteness of the system and the nonlinearity of the dispersion relation at small wavenumbers.

We summarize the findings of this section: Hawking radiation with a thermal spectrum is emitted from a black hole horizon on an ion ring, even for a finite system with a logarithmically diverging group velocity at low wavenumbers due to long range interactions (see Fig. \ref{dispersion}).

\subsection{Correlations}
\label{sec.correlations}
For an experimental proof of Hawking radiation on ion rings, we propose to observe the emission of Hawking radiation following the creation of a black hole horizon (see Sec. \ref{Experimental Parameters}). We propose to measure the emitted phonons or the emerging correlations between the subsonic and the supersonic region (see \cite{Carusotto08}). The latter is discussed in Sec. \ref{Review Correlations}, in this section we are presenting simulation results for ion rings on the emerging correlations. We compare our results to the analytical findings for a continuum system with linear dispersion relation derived in Eq. \eqref{momentum.corr} \cite{Carusotto08}. In the quantum regime the emergent cross-correlations display the generation of entanglement. We study its properties, especially to analyze the crossover from the quantum to the classical Hawking effect.

\subsubsection{Discrete System}
\label{Discrete System}
\begin{figure}[t]
\begin{center}
\includegraphics[width=75mm,height=57mm,angle=0]{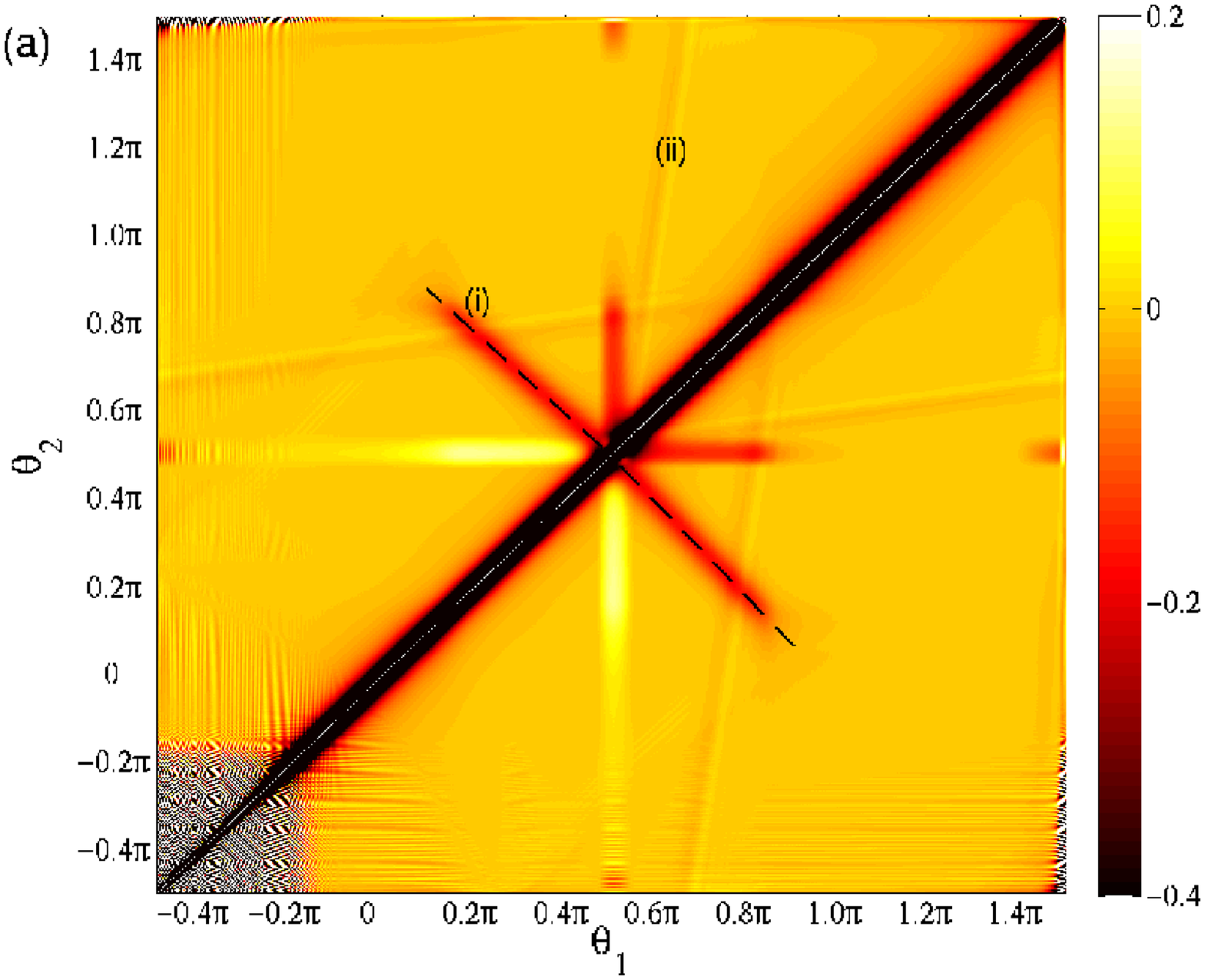}
\includegraphics[width=75mm,height=57mm,angle=0]{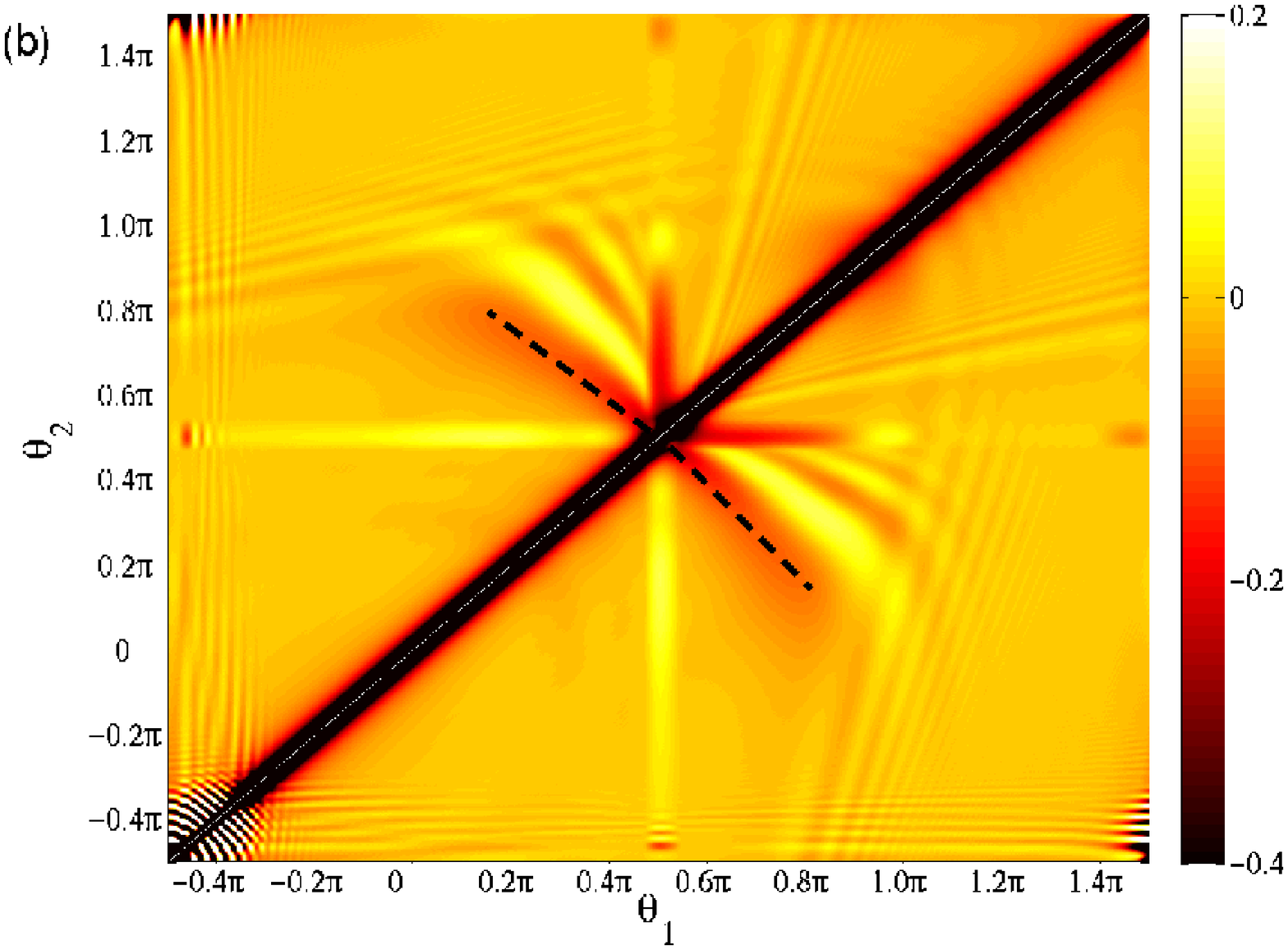}
\caption{Momentum-momentum correlations $C_{ij}(t)$ at time $t=0.5T$ with real space lab frame positions. Starting from homogeneously spaced ions with temperature $T_0=0$ a black hole region is created in the small time interval $\tau=0.05T$. The dashed lines point to the signature corresponding to the emission of pairs of Hawking phonons. We consider $N=1000$ ions, and $\sigma v_\text{min}T=2\pi\cdot 0.25$. (a) Nearest-neighbor interactions with $e^2/4\pi\epsilon_0=\frac{1.127}{2N}\frac{mL^3}{T^2}$; (b) Full Coulomb interactions with $e^2/4\pi\epsilon_0=\frac{0.2453}{2N}\frac{mL^3}{T^2}$ \cite{Horstmann10} (see \ref{section velocity profile}.}
\label{densitydensity}
\end{center}
\end{figure}


We propose the experiment to start from the ground/thermal state of the excitation around homogeneously spaced ions at rest with temperature $T_0$. Then the system is accelerated with a constant force which does not change the quantum state of the system defined relative to the equilibrium positions. Subsequently, a supersonic region is created in the small time interval $\tau$. This is done by reducing the subsonic fluid velocity $v_\text{min}$ in a Gaussian way, while leaving the average rotation velocity constant (see \ref{section velocity profile} and Eq. \eqref{Reduce vmin}). We reduce excitations created at the white hole horizon at $\theta/2\pi\approx 1-\sigma v_\text{min} T$ by a wider transition region at this horizon. In an experiment the magnitude and velocity of these excitations can be watched through careful measurements.

We are analyzing the momentum-momentum correlations
\begin{equation}
\label{momentum-momentum correlations simulation}
 C_{ij}={\langle \delta \hat{p}_i \delta \hat{p}_j\rangle}\cdot T/(\hbar m).
\end{equation}
In the continuum limit these momenta correspond to the time derivative of the scalar field $\hat{\Phi}$
\begin{equation}
 \delta \hat{p}_i \sim \kla{\partial_t+v(\theta)\partial_\theta}\hat\Phi(\theta,t).
\end{equation}
The dynamics of the momentum-momentum correlations are given by the dynamics of the covariance matrix (see Eq. \eqref{Gamma_dynamics}). In Sec. \ref{Section Measurement} we are explaining how these correlations $C_{ij}$ can be measured in an experiment. 

\begin{figure}[t]
\begin{center}
\includegraphics[width=75mm,height=57mm,angle=0]{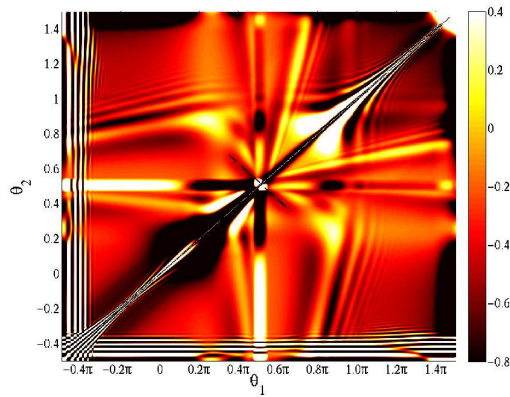}
\caption{Momentum-momentum correlations $C_{ij}(t)$ for initial temperature $T_0=102T_\text{H}$ at time $t=0.5T$ with real space lab frame positions. We consider $N=1000$ ions, $\sigma v_\text{min}T=2\pi\cdot 0.25$, $e^2/4\pi\epsilon_0=\frac{0.2453}{2N}\frac{mL^3}{T^2}$, and $\tau=0.05T$. Full Coulomb interactions are considered (compare with Fig. \ref{densitydensity}).}
\label{densitydensityfullhot}
\end{center}
\end{figure}

Fig. \ref{densitydensity} shows the simulation results for $C_{ij}$. It displays a fixed time after the black hole formation starting from the ground state, i.e., $T_0=0$. Correlations between the inside and the outside of the black hole are created close to the black hole horizon and are moving away from it as expected. As stated earlier these correlations correspond to the pairs of Hawking particles. We interpret their pure existence as a signature for Hawking radiation. Fig. \ref{densitydensity}(a) shows the simulation for interactions between neighboring ions only, Fig. \ref{densitydensity}(b) for long range Coulomb interactions. The correlations behave similarly in both cases (see discussion below). They can still be observed for initial temperatures two orders of magnitude above the Hawking temperature \cite{Carusotto08} as shown in Fig. \ref{densitydensityfullhot}. In Sec. \ref{Entanglement Generation} we find that the Hawking effect is still quantum at such initial temperatures (see Fig. \ref{logneg}). In contrast to the entanglement, the cross-correlations actually remain present at arbitrarily large initial temperatures \cite{Recati09} (note $\Gamma\propto T_0$ for $k_\text{B}T_0\gg \hbar\omega_\text{k}$, see Eq. \eqref{thermal state}).

The most significant correlation signal (i), i.e. the line of negative cross-correlations, demonstrates a basic property of Hawking radiation, it corresponds to two upstream phonons, one inside and one outside the black hole. An additional correlation feature (ii) is fully inside the black hole, corresponding to a pair of upstream and downstream phonons inside the black hole. The features (i) and (ii) have already been reported for analog black holes in a BEC (see \cite{Carusotto08}). The propagation velocity of these correlations $c(\theta)\pm v(\theta)$ depends on the group velocity, i.e. the dispersion relation, and the ion velocity. The angle of the cross-correlation signature (i) is determined by the ratio of the phonon velocities inside and outside of the black hole $\kla{c(\theta)-v(\theta)}/ \kla{c(\theta')-v(\theta')}$. The dashed lines in Fig. \ref{densitydensity} show the predictions for the direction of the cross-correlation signal (see Sec. \ref{Review Correlations}).

In contrast to Fig. \ref{densitydensity}(a), which is very similar to black hole analogues in a BEC \cite{Carusotto08}, the correlation plot for full Coulomb interactions in Fig. \ref{densitydensity}(b) displays a more complicated structure. The cross-correlation signal, is broader and there are additional lines of oscillating correlations. With higher resolution and/or for larger Hawking temperatures these effects can also be observed for nearest-neighbor interactions. We attribute these changes to the more complicated and sublinear dispersion relation.

Even though the precise determination of the correlations for a nonlinear dispersion relation is a rather complicated issue (see, e.g., \cite{correlations}), we may obtain a qualitative understanding by means of the following simple picture. Since group and phase velocity nearly coincide at small wavenumbers, the correlations are created near the horizon as a nice $1/\cosh^2$-pulse as in the case of a linear dispersion relation (see Eq. \eqref{momentum.corr}).

\begin{figure}[t]
\begin{center}
\includegraphics[width=100mm,angle=0]{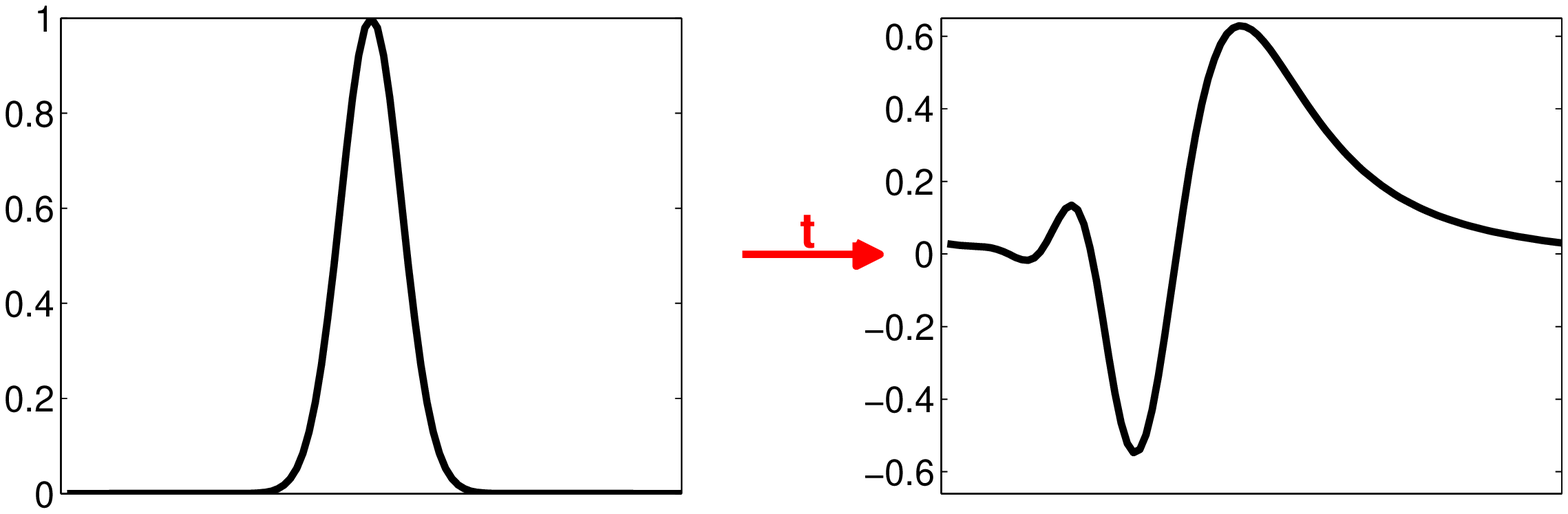}
\caption{Schematic depiction of the pulse propagation on an ion ring. The left graph depicts an initial Gaussian pulse profile, the right graph the dispersed pulse after time $t$. Trails of oscillations are following the dispersed pulse.}
\label{pulse dispersion}
\end{center}
\end{figure}

However, as this pulse propagates away from the horizon, the non-linear dispersion relation deforms it. For a fixed $t',\theta'$, the two-point function $\langle \delta \hat{p}(t,\theta) \delta \hat{p}(t',\theta')\rangle$
obeys the same wave equation as $\delta \hat{p}(t,\theta)$ itself. In a homogeneous region of a stationary spacetime this is the same as for the field $\Psi$ (see Eq. \eqref{field equation}). 
\begin{equation}
\left\{\klab{\partial_t+\partial_\theta v}\klab{\partial_t+v\partial_\theta}-\klab{iD\kla{-i\partial_\theta}}^2 \right\}
\langle \delta \hat{p}(t,\theta) \delta \hat{p}(t',\theta')\rangle=0.
\end{equation}
Since modes with larger $k$ propagate slower than those with smaller $k$, a pulse with an initial $1/\cosh^2$-shape will be deformed during the time evolution similar to Fig. \ref{pulse dispersion}. The main pulse (global maximum) becomes broader and oscillations develop, trailing the main pulse (local minima and maxima), which are caused by the slower modes with short wavelengths. This deformation applies to the outgoing Hawking radiation and the 
infalling partners in the same way as both are moving upstream. In a finite system the deviations from a linear dispersion are significantly larger for full Coulomb interactions than for nearest-neighbor interactions. As a result, this simple picture explains the difference between Figs. \ref{densitydensity}(a) and \ref{densitydensity}(b). 

\subsubsection{Comparison with Continuum System}
In Sec. \ref{Review Correlations} we have calculated and analyzed the cross-correlation signal for a continuum system with strictly linear dispersion relation. We observe these correlations also for our discrete system with sublinear dispersion relation as shown in Fig. \ref{densitydensity}. In this Section we are quantitatively comparing the simulation results for ion rings with the analytical results for a continuous system, focussing on the peak magnitude of the cross-correlations (see Eq. \eqref{momentum.corr}).

\begin{figure}[t]
\begin{center}
\includegraphics[width=75mm,height=57mm,angle=0]{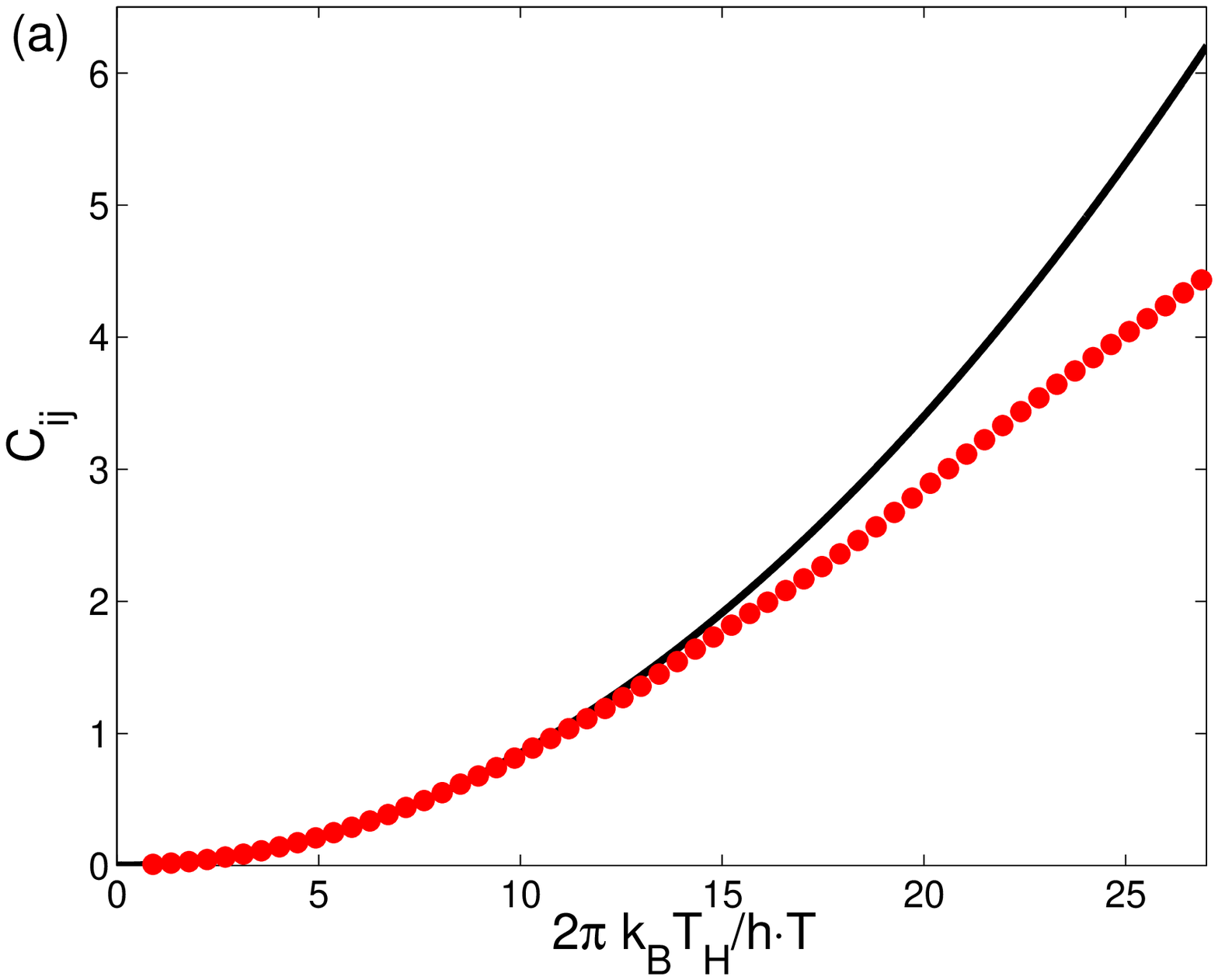}
\includegraphics[width=75mm,height=57mm,angle=0]{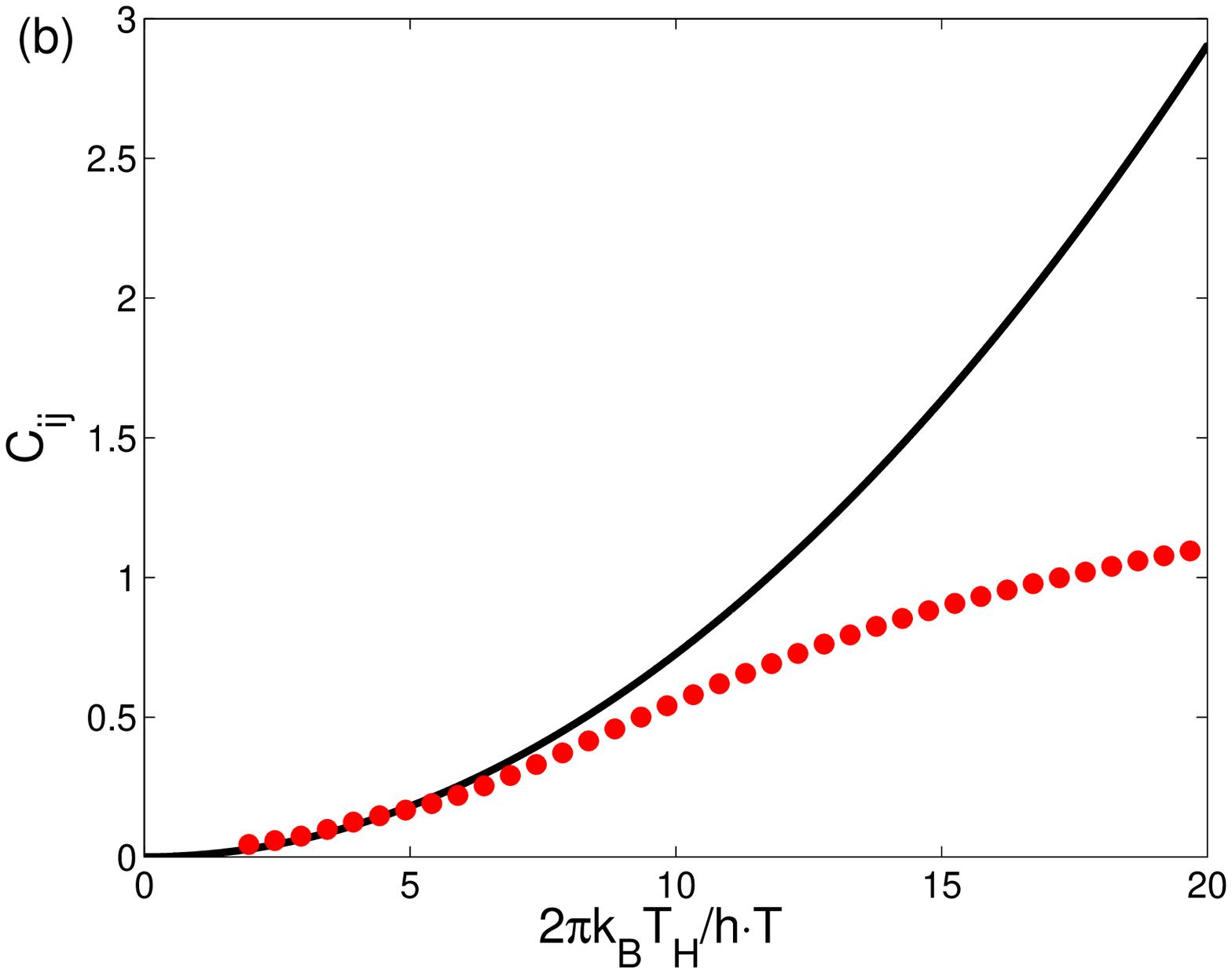}
\caption{Peak Height $C_{ij}(t)$ of the cross-correlation signature in the momentum-momentum correlations (see Fig. \ref{densitydensity}) as a function of the Hawking temperature $T_\text{H}$. The dots are simulation results for a discrete system with nearest-neighbor interactions only, $N=1000$ ions, $\sigma v_\text{min}T=2\pi\cdot 0.25$, and $\tau=0.05T$. $\gamma_1$ is varied to get different Hawking temperatures $T_\text{H}$. $h=2\pi\hbar$ is Planck's constant. (a) Nearest-neighbor interactions with $e^2/4\pi\epsilon_0=\frac{1.127}{2N}\frac{mL^3}{T^2}$ and $0.00\bar{3}<\gamma_1<0.1$; (b) Full Coulomb interactions with $e^2/4\pi\epsilon_0=\frac{0.2453}{2N}\frac{mL^3}{T^2}$, and $0.005<\gamma_1<0.1$ (see \ref{section velocity profile}.}
\label{peakheight}
\end{center}
\end{figure}

We perform this comparison for varying Hawking temperatures. It is tuned by changing the widths of the horizon region $\gamma_1$ (see Eq. \eqref{metricg}) keeping constant the other parameters. The comparison is shown for nearest-neighbor interactions in Fig. \ref{peakheight}(a) and for full Coulomb interactions in Fig. \ref{peakheight}(b), where the dots represent the simulated peak magnitude for the ion ring and the curve represents the analytic peak magnitude for the continuum system with a linear dispersion relation.

For nearest-neighbor interactions, the results agree very well at small Hawking temperatures; for full Coulomb interactions, the agreement is quite good at moderate Hawking temperatures. The deviation at large Hawking temperatures is caused by the discreteness of the system. For large Hawking temperatures the horizon region, which almost completely determines the properties of Hawking radiation, is small and contains only a few particles. Then the discreteness of the system becomes relevant. In the limit of very large Hawking temperatures the peak height should mainly depend on the lattice spacing \cite{Recati09}.

\subsubsection{Entanglement Generation}
\label{Entanglement Generation}
In this section we are discussing the creation of entanglement between the inside and the outside of a black hole following its creation. Entanglement is unique to quantum processes. Therefore, the existence of entanglement proves that one can observe the quantum version of the Hawking effect with our proposal. In contrast, correlations between the inside and the outside of the black hole are present both for initial thermal states $T_0\gg T_\text{H}$ in the \emph{classical} regime and for initial \emph{quantum} states $T_0=0$ \cite{Recati09}. So we analyze the crossover between classical and quantum Hawking radiation (stimulated versus spontaneous emission). Furthermore, our analysis emphasizes the importance we assigned to the cross-correlations. The emerging entanglement can be measured on two routes, either by measuring the covariance matrix through a measurement of correlation in the ion displacements (see Sec. \ref{Section Measurement}) or by swapping the entanglement from the motional to the internal degrees of freedom of the ions \cite{Retzker05}.

We are now briefly introducing the relevant entanglement measures before we present numerical results for the ion system. The covariance matrix $\Gamma$ (see Eq. \eqref{Gamma}) can be calculated (see \ref{equations of motion}) and measured (see Sec. \ref{Experimental Parameters}) for the ion system in harmonic approximation (see Eq. \eqref{Hamiltonian}). It gives access to two entanglement measures: the entropy of entanglement and the logarithmic negativity.

\begin{figure}[t]
\begin{center}
\includegraphics[width=75mm,height=57mm,angle=0]{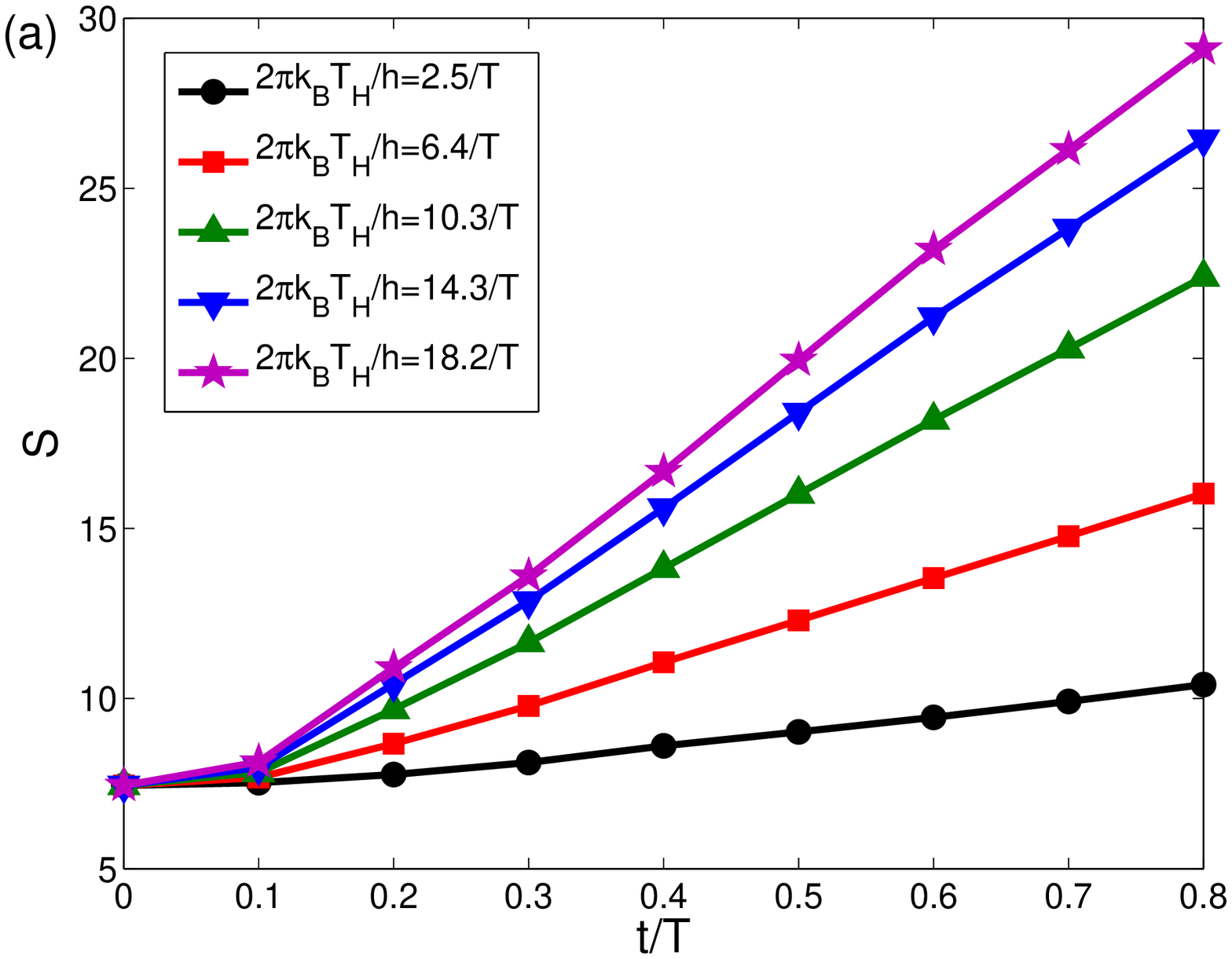}
\includegraphics[width=75mm,height=57mm,angle=0]{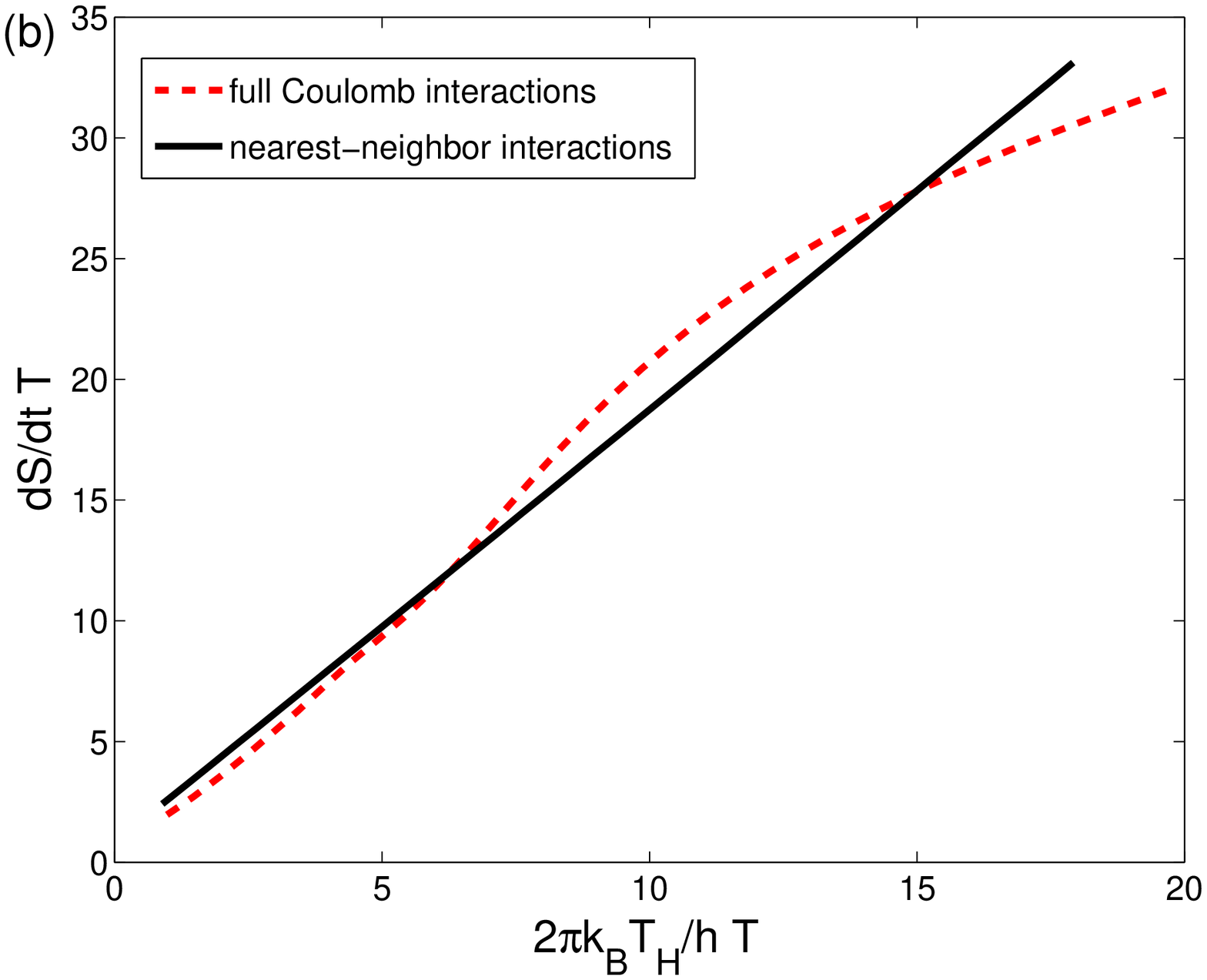}
\caption{(a) Time evolution of the entropy of entanglement ${\bf S}(\Gamma)$ for different Hawking temperatures $T_\text{H}$ with full Coulomb interactions (same parameters as Fig. \ref{peakheight}(b)). (b) Rate of increase of the entropy of entanglement $d{\bf S}(\Gamma)/dt$ as a function of the Hawking temperature $T_\text{H}$. Full Coulomb interactions (red dashed line, same parameters as Fig. \ref{peakheight}(b)), and nearest-neighbor interactions (black straight line, same parameters as Fig. \ref{peakheight}(a)). The linear increase in entanglement is caused by the constant emission of entangled Hawking phonons.}
\label{eoe}
\end{center}
\end{figure}

We consider a system with density matrix $\rho$ divided into subsystems A and B. The entropy of entanglement is defined for a bipartite pure state. It is the Von-Neumann entropy of the reduced density matrix $\rho_A$ of one subsystem $A$ \cite{Bennet96}
\begin{equation}
 {\bf S}(\rho)=-\rm{tr}\kla{\rho_A \log_2 \rho_A}.
\end{equation}
The state $\rho$ is a product state for ${\bf S}(\rho)=0$. As an entanglement measure $\bf S$ cannot increase under local operations and classical communications (LOCC). For pure Gaussian states ${\bf S}(\rho)$ can efficiently be calculated from the covariance matrix $\Gamma$ \cite{Vidal02, Reznik03, Wolf04}. It is given by
\begin{equation}
{\bf S}(\Gamma)=\sum_{n=1}^N \kla{\lambda_n^2\log_2\lambda_n^2-(\lambda_n^2-1)\log_2\kla{\lambda_n^2-1}}
\end{equation}
with the symplectic eigenvalues $\lambda_n,\text{ } n=1,\dots,N$ of the covariance matrix $\Gamma$. They are the eigenvalues of $i\sigma\Gamma$ with the symplectic matrix 
\begin{equation}
\sigma=\bigoplus_{n=1}^N\begin{pmatrix} 0& 1\\ -1&0 \end{pmatrix},
\end{equation}
which exchanges position and momentum of each mode.

We can also use the logarithmic negativity, which is an entanglement monotone, i.e., it does not decrease under LOCC \cite{Vidal02}. In contrast to the entropy of entanglement, the logarithmic negativity can be calculated efficiently from the covariance matrix even for mixed states. It is defined as the logarithm of the trace-norm of the partial transpose of the density matrix
\begin{equation}
{\bf N}(\rho)=\log_2 \lVert\rho^{T_a}\rVert
\end{equation}
with $\lVert M \rVert=\sqrt{M^\dagger M}$. The covariance matrix $\Gamma^{T_a}$ of the partial transpose of $\rho$, i.e., $\rho^{T_a}$, follows from the covariance matrix $\Gamma$ of $\rho$ by multiplying with $-1$ all matrix entries of $\Gamma$ which contain exactly one momentum operator of subsystem A. Let $\widetilde{\lambda_k},\text{ } k=1,\dots,N$ denote the symplectic spectrum of $\Gamma^{T_a}$. Then the logarithmic negativity is
\begin{equation}
 {\bf N}(\Gamma)=-\sum_{\substack{n=1\\2\widetilde{\lambda_n}<1}}^N \log_2 \kla{2\widetilde{\lambda_n}}.
\end{equation}
A vanishing logarithmic negativity does not mean that the system is not entangled, but it means that such systems cannot be purified to maximally entangled states.


Now, we analyze the entanglement properties for black holes on ion rings. Using the entropy of entanglement we study how the entanglement is generated between the inside and the outside of a black hole starting from the ground state ($T_0=0$) at initial times. The entropy of a black hole is more fundamental than the logarithmic negativity, but it is only defined for pure states. Thus, we have to compare the whole supersonic with the whole subsonic region. On a ring this means that one cannot determine at which horizon the entanglement is created. 

In Fig. \ref{eoe}(a) we find a linear increase of the entropy of entanglement in time after an initial period ($t\sim 0.2T$ in our case). This linear increase corresponds to the constant emission of Hawking radiation from a black hole. We have plotted the rate of this increase in Fig. \ref{eoe}(b). We find for nearest-neighbor interactions a linear dependence of the entropy of entanglement on the Hawking temperature of the black hole. For full Coulomb interactions we find a similar behavior. 

\begin{figure}[t]
\begin{center}
\includegraphics[width=80mm,height=60mm,angle=0]{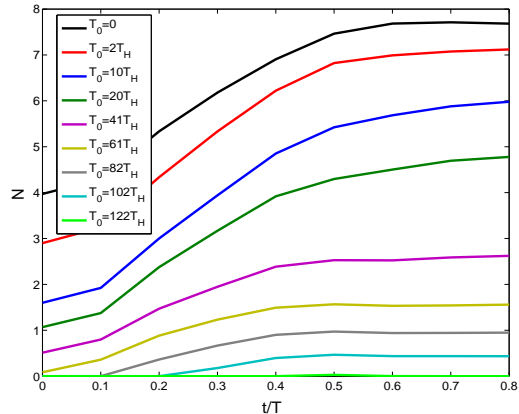}
\caption{Time evolution of the logarithmic negativity $\bf N$ of a region with $0.2N$ ions ($N=1000$) adjacent to the horizon for different initial temperatures $T_0$. The linear increase in entanglement is caused by the constant emission of entangled Hawking phonons. Full Coulomb interactions considered (same parameters as Fig. \ref{densitydensity}(b)).}
\label{logneg}
\end{center}
\end{figure}

The logarithmic negativity gives access to the entanglement developing for systems with finite initial temperatures $T_0>0$. The entanglement between two small regions adjacent to the black hole is presented in Fig. \ref{logneg}. The initial entanglement between the regions depends on the system temperature: For quantum systems at $T_0=0$ entanglement is present, for classical systems $T_0\rightarrow\infty$ it cannot be detected. The entanglement is increasing linearly in time for sufficiently small $T_0$. This behavior is already described above for the entropy of entanglement (see Fig. \ref{eoe}(a)).

However, the logarithmic negativity allows further observations. For initial temperatures $T_0$ more than two orders of magnitude above the Hawking temperature $T_\text{H}$, no entanglement generation is visible in the logarithmic negativity. In contrast, the cross-correlation signal remains present for arbitrarily high initial temperatures. We observe here the transition from the \emph{quantum} to the \emph{classical} Hawking effect. We can conclude that for the initial temperatures $T_0< T_0^\text{c}\approx 100 T_\text{H}$ ($N=1000$) the observed Hawking effect is quantum, whereas one would naively expect this transition at $T_0^\text{c}\approx T_\text{H}$. $T_0^\text{c}/T_\text{H}$ is increasing with the number of ions, taking, for example, $N=100$ ions and $e^2/4\pi\epsilon_0=0.6/(2N)\cdot (mL^3)/T^2)$ we find $T_0^\text{c}\approx 24T_\text{H}$. $T_0^\text{c}/T_\text{H}$ is approximately proportional to the largest mode frequency of the system, which is sublinear in in $N$. 

This might be understood with the following argument: As described in Sec. \ref{Review Scattering} Hawking radiation emerges from large wavenumbers before being emitted at small wavenumbers. The frequency related to these large wavenumbers is about $N$-times (number of ions) higher than the smallest frequency in the system. Along this line, the Hawking effect remains quantum for initial temperatures comparable to these highest frequencies of the system at large wavenumbers.

In Fig. \ref{logneg} a saturation in the logarithmic negativity at later times is observed. This is in agreement with the fact that Hawking radiation is constantly emitted. After fully penetrating the small regions adjacent to the horizon, no additional entanglement can develop between these regions.

\section{Experimental Realization}
\label{Experimental Parameters}
So far we have studied the appearance of Hawking radiation on ion rings. We have demonstrated that the emitted radiation has a thermal spectrum with a geometrically justified Hawking temperature (see Sec. \ref{sec.scattering}), and that it is created in pairs (see Sec. \ref{sec.correlations}). We complete our analysis in this section by describing an experimental setup which will allow for the observation of Hawking radiation on an ion ring.

First, we present a suitable parameter regime for an experiment. The main condition on the experimental parameters is that the ion velocity must be approximately equal to the phonon velocity. This condition leads to the requirement $e^2/4\pi\epsilon_0\approx 0.25/\kla{2N} mL^3/T^2$. For $N=1000$ singly charged $^9Be$ ions with an average spacing of $L/N=2\mu m$ the rotation frequency of the ions would be $\omega_\text{rot}=2\pi\times 120kHz$. $\omega_\text{rot}$ represents the smallest mode frequency of the system. For $v_\text{min}=(2\pi\times 0.8\bar{3})/T$, $\gamma_1=0.02$, and $\sigma v_\text{min}T=2\pi\cdot 0.25$ the Hawking temperature is $k_\text{B}T_\text{H}/\hbar\approx 5/T\approx 2\pi\times 95kHz$. If the initial temperature is two orders of magnitude above the Hawking temperature $T_0\lesssim 100T_\text{H}$, we show explicitly in Sec. \ref{Discrete System} that the cross-correlation signature of Hawking radiation remains present (see Fig. \ref{densitydensityfullhot}) and we find in Sec. \ref{Entanglement Generation} that the Hawking radiation remains a quantum effect. Thus, it is not necessary to perform ground state cooling of all vibrational modes of the ions in an experiment.

Note that it has been demonstrated long ago how to trap ions in quadrupole ring traps \cite{Drees64} and measure their arrangement \cite{Walther92}. The ideas of these experiments can be combined with modern cooling techniques applied to ions in linear Paul traps or in surface traps \cite{Morigi2003, Wineland09}. Thus, the proposed experiment will allow to measure signatures of Hawking radiation for acoustic black holes with parameters and temperatures which can be reached in current experiments.

The general idea of the actual measurement process is the following. The Hawking effect is encoded in the motional degrees of freedom of the ions, which are described in this paper with the ion displacements $\delta\theta_i$ and can be viewed as phonic modes. A different degree of freedom for ions is their internal state, here we address two hyperfine states of the ions. Lasers couple the motional degrees of freedom to the two relevant internal states. In this way, the information on the Hawking effect is transfered to the internal states. The occupation of the internal states can be read out by fluorescence imaging.

In Sec. \ref{Section Measurement} we first propose a measurement sequence for the cross-correlation signature. Then we discuss a proposal to directly measure the Hawking phonons in Sec. \ref{Direct Measurement}. The proposal in Sec. \ref{Section Measurement} allows to determine any part of the covariance matrix (see Eq. \eqref{Gamma}). Thus, it can be used to determine the emerging entanglement between the inside and the outside of the black hole (see Sec. \ref{Entanglement Generation}). Note that it has been proposed earlier how to detect entanglement in the motional degrees of ions \cite{Retzker05}. The basic mechanism of all proposals is the coupling of the ion displacements to their internal levels with lasers.

\subsection{Measurement of Ion Displacements}
\label{Section Measurement}
In this section we discuss how the cross-correlation signal analyzed in Sec. \ref{Discrete System} can be detected in an experiment. We propose to measure correlations in the ion displacements by coupling the motional degrees of freedom of the ions to their internal states. 

First, we explain how to relate the momentum-momentum correlations we discussed (see Eq. \eqref{momentum-momentum correlations simulation}) to experimentally accessible correlations in the ion displacement and analyze how accurately the latter should be measured. To this aim we rewrite Eq. \eqref{momentum_angle} for a continuum system in a region of constant flow as
\begin{equation}
\delta\hat p(\theta)=\frac{L}{2\pi}mc(\theta)n(\theta)\frac{\partial_\theta\hat\Phi(\theta,t)}{n(\theta)}.
\end{equation}
We can thus measure the momentum-momentum correlations by spatial derivatives of the ion displacements
\begin{equation}
\left< \kla{\delta\hat\theta_i-\delta\hat\theta_{i+\Delta}}\kla{\delta\hat\theta_j-\delta\hat\theta_{j+\Delta}} \right>
=\kla{\frac{2\pi}{mL}}^2\frac{\Delta^2}{n_i n_j c_i c_j}\langle \delta\hat p_i\delta\hat p_j\rangle.
\end{equation}
The analysis shown in Fig. \ref{peakheight}(b) confirms that we can use Eq. \eqref{momentum.corr} to get the order of magnitude of the momentum-momentum correlations for a finite ion ring. Thus, we can estimate the magnitude of the cross-correlation signal as
\begin{equation}
\left< \kla{\delta\hat\theta_i-\delta\hat\theta_{i+\Delta}}\kla{\delta\hat\theta_j-\delta\hat\theta_{j+\Delta}} \right>
\approx \kla{\frac{2\pi}{L}}^2\frac{\hbar T}{m}\frac{\pi^3\Delta^2}{N^3}\frac{\kla{k_\text{B} T_\text{H}/\hbar}^2}{(c_i-v_i)(c_j-v_j)},
\end{equation}
where we used $n_i\approx n_j\approx N/(2\pi)$ and $c_i\approx c_j\approx (2\pi)/T$. Therefore, the angle-angle correlations must be detected in an experiment with the accuracy
\begin{equation}
\epsilon :=\Delta \langle \delta\hat\theta_i\delta\hat\theta_j \rangle = \kla{\frac{2\pi}{L}}^2\frac{\hbar T}{m}\frac{\pi^3\Delta^2}{4N^3}\frac{\kla{k_\text{B} T_\text{H}/\hbar}^2}{(c_i-v_i)(c_j-v_j)}.
\end{equation}
In the following we are describing a setup to detect the angle-angle correlations of the ions with sufficient accuracy. We propose to illuminate the ion ring at two positions. The lasers should be focussed on one ion inside the supersonic and one ion inside the subsonic region on the ring. At each position a laser beam couples two internal levels $\ket g$ and $\ket e$ of the ions with the transition energy $\omega_I$. The lasers should fullfill the resonance conditions
\begin{equation}
 \omega=\omega_I+kv\frac{L}{2\pi}\\
\end{equation}
for their frequencies and wavenumbers, which takes into account the Doppler shift $kv$. The ion traverses the pulse beam in the time $T/N$ which is much shorter than the time scale $T=2\pi/\omega_\text{rot}$ of phonons at small wavenumbers. Thus, we can neglect the ion motion for the further analysis. After going to the frame rotating with $\omega_0$ and applying the rotating wave approximation the dipolar coupling Hamiltonian \cite{Windeland98} of one illuminated ion becomes
\begin{equation}
 \mathcal{H}_{dip}=\hbar\Omega\kla{\sigma^+e^{\frac{ ikL}{2\pi}\delta\hat \theta}+H.c.},
\end{equation}
where $\Omega$ is the Rabi frequency of the laser transition. Note that the ions remain in the Lamb-Dicke limit
\begin{equation}
\label{Lamb_dicke}
 \sqrt{\frac{\hbar}{mN\omega_\text{rot}}}k\ll 2\pi
\end{equation}
during the experiment, i.e. $kL\delta\hat\theta\ll 4\pi^2$.

We propose to prepare the internal state of each ion in the superposition $(\ket g + \ket e)/\sqrt{2}$ before the experiment. After the creation of the black hole and the illumination with the lasers the probability that the two measured ions are in the states $\ket{gg}$ or $\ket{ee}$ 
\begin{equation}
P\kla{\delta\hat \theta_1,\delta\hat \theta_2}=\frac{1}{2}+\kla{\frac{L}{2\pi}}^2\sin^2\kla{2\Omega t}\frac{k^2}{2}\langle \delta\hat \theta_1\delta\hat \theta_2 \rangle
\end{equation}
is measured through a repetition of the experiment. We propose to use the Rabi frequency $2\Omega t=\pi/2$. If the measurement is repeated $M$ times, the standard deviation of the average number of binomially distributed events $P$ is
\begin{equation}
\Delta P=\sqrt{\frac{P(1-P)}{M}}\approx \frac{1}{2\sqrt{M}}.
\end{equation}
It should be smaller than the required accuracy of the signal size and thus
\begin{equation}
M>\kla{\frac{2\pi}{L}}^4 k^{-4} \epsilon^{-2}\approx\kla{\frac{m\omega_\text{rot}}{\hbar k}}^2\klab{\frac{2N^3}{\Delta ^2\pi^4}\frac{(c_i-v_i)(c_j-v_j)}{(k_\text{B}T_\text{H}/\hbar)^2}}^2
\end{equation}
measurements are necessary. 

\begin{figure}[t]
\begin{center}
\includegraphics[width=75mm,height=57mm,angle=0]{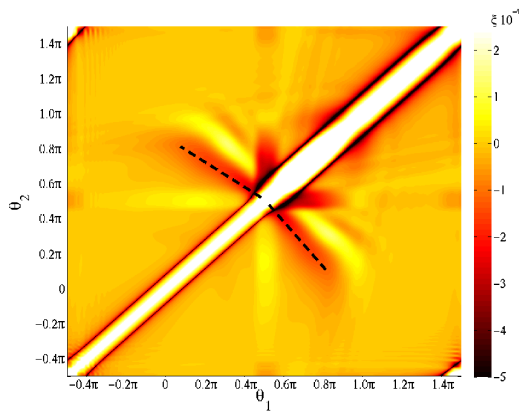}
\caption{Correlations $\langle (\delta\hat\theta_{i}-\delta\hat\theta_{i+50})(\delta\hat\theta_{j}-\delta\hat\theta_{j+50}) \rangle\cdot m/(\hbar T)\cdot L^2/(2\pi)^2$ at time $t=0.6T$ with real space lab frame positions. These correlations are analougous to the momentum-momentum correlations shown in Fig. \ref{densitydensity}b. We consider $N=1000$ ions, $\sigma v_\text{min}T=2\pi\cdot 0.25$, $e^2/4\pi\epsilon_0=\frac{0.2453}{2N}\frac{mL^3}{T^2}$, and $\tau=0.05T$. Full Coulomb interactions are considered.}
\label{densitydensityfullmeasurement}
\end{center}
\end{figure}

We are now calculating $M$ for the example of $N=1000$ $^9Be$ ions, discussed in the beginning of this Sec. \ref{Experimental Parameters}. We propose to use the $\lambda=313nm$ transition in $^9Be$ \cite{Monroe95}. We have checked that the cross-correlations remain clearly visible for the resolution $\Delta=50$ (see Fig. \ref{densitydensityfullmeasurement}). In this case $M>1.1\cdot 10^{5}$ measurements are required. This basic measurement proposal can certainly be improved by employing additional techniques, e.g., using interferences with additional lasers. But we do not propose further experimental setups here, since these should be adapted to specific implementations of our proposal.

The accuracy of the classical equilibrium positions of the ions must satisfy $|\Delta\theta_i^0|^2<\sqrt{\epsilon}$, such that it does not influence the cross-correlation signal. Inaccurate equilibrium positions can be caused by inaccurate external forces (see \ref{system detail}) during the black hole formation. Assuming that the external force is $\widetilde F_i(t)=(1+\gamma)F_i(t)$ (see Eq. \eqref{eq.external force}) with $\gamma\ll 1$, the deviation in the classical positions fullfills the equation
\begin{equation}
\label{DGLdelta}
 \frac{L}{2\pi} m\frac{d^2\Delta\theta_i}{dt^2} = \gamma F_i(t).
\end{equation}
Integrating Eg. \ref{DGLdelta}, we can estimate
\begin{eqnarray}
\label{eq.forces.accuracy}
\frac{|\Delta\theta_i|}{\gamma}&<&\left|\theta_i^0(\tau)-\theta_i^0(0)-\frac{d\theta_i^0(0)}{dt}\tau\right|+\frac{\tau^2}{2}\frac{2\pi}{L} \frac{|F_i^\text{c}|}{m}\nonumber\\
&\approx&|\sigma v_\text{min} T -2\pi\sigma|+
\frac{\tau^2}{2}\frac{2\pi}{L} \frac{e^2}{4\pi\epsilon_0m}\frac{N^2}{L^2}(v_\text{max}-v_\text{min})\frac{\pi^2}{6} 
\end{eqnarray}
For $\tau=0.05T$ we get $|\Delta\theta_i|<2\pi\cdot 0.25\gamma$, thus the accuracy $\gamma<\sqrt{\epsilon}/(2\pi\cdot 0.25)\approx 5 \cdot10^{-6}$ is required for the external forces.

\subsection{Measurement of Hawking Phonons}
\label{Direct Measurement}
The measurement scheme described in the previous section requires the accurate control of the ion acceleration during the creation of the black hole (see Eq. \eqref{eq.forces.accuracy}). Here we propose a scheme to avoid this difficulty (see also \cite{detect}). If the number of ions in the supersonic region $(1-2\sigma)N$ is small, the displacement of the ions due to the creation of the black hole will be small compared to the equilibrium motion of the ions. Then the creation of the black hole is adiabatic. In this case the measurement of cross-correlations is not possible. Instead, we describe in this subsection how to directly measure the emitted Hawking phonons outside of the black hole.

The following setup is studied: After the black hole formation in the small time interval $\tau$, Hawking phonons are emitted at the black hole horizon. We propose to detect these phonons by coupling the oscillation of the ions to their internal state with a laser. The laser drives a transition between two internal states $\ket{a}$ and $\ket{b}$ of the ions, which are prepared in the state $\ket{a}$. It illuminates $\widetilde N$ ions outside of the black hole and is following the motion of these ions for the time interval $t_m$. Since $t_m$ determines the spectral width of the laser, it must be large compared to the inverse Hawking temperature $t_m\gg \hbar/(k_\text{B}T_\text{H})$. The laser frequency $\omega$ should fullfill the resonance condition
\begin{equation}
 \omega=\omega_I+\omega_{p_0}+kv\frac{L}{2\pi},\\
\end{equation}
which takes into account the transition frequency of the ions $\omega_I$, the Doppler shift $kv$, and the relevant phonon frequency $\omega_{p_0}$. After going to the frame rotating with $\omega_I$ and $\omega_{p_0}$ and applying the rotating wave approximation the dipolar coupling Hamiltonian \cite{Windeland98} in the Lamb-Dicke limit (see Eq. \eqref{Lamb_dicke}) becomes
\begin{equation}
\mathcal{H}_{dip}=\sum_{j,p} \kla{\frac{\hbar\Omega_p}{2}\sigma_j^+a_p e^{i(p+k)Lj/N}+H.c.}+
\sum_p \hbar\kla{\omega_p-\omega_{p_0}} a_p^\dagger a_p,
\end{equation}
where the sums extend over the illuminated ions $j$ and the relevant phonon modes $p$. $k$ denotes the wavenumber of the laser and 
\begin{equation}
\Omega_p=-i\Omega k\sqrt{\frac{2\hbar}{m\omega_p}}
\end{equation}
are the effective Rabi frequencies with the bare Rabi frequency $\Omega$. We describe here the coupling of the phonon modes to a sideband transition. In the limit of a laser pulse with small spectral width $\widetilde{N}\ll N t_m/T$ we can introduce the spin $\Sigma=\widetilde{N}/2$ operator $\Sigma^+=\sum_j \exp\kla{-ikLj/N}\sigma_j^+/2$ and transform it into a bosonic field $b^\dagger$ with the Holstein-Primakoff transformation
\begin{equation} 
\Sigma^+=b^\dagger\sqrt{2\Sigma-b^\dagger b}\approx b^\dagger\sqrt{\widetilde{N}} 
\end{equation}
For $\langle b^\dagger b\rangle \approx \langle a_{p}^\dagger a_{p}\rangle \lesssim k_\text{B}T_\text{H}/(\hbar\omega_\text{rot})\ll\widetilde{N}$ (see Eq. \eqref{time_evolution_b}), we can use $\Sigma^+\sim \sqrt{\widetilde{N}}b^\dagger$. The expectation value $\langle b^\dagger b\rangle$ gives the number of ions in the state $\ket{b}$. So the coupling of the relevant phonon modes to the illuminated ions can be described by the Hamiltonian
\begin{equation}
\mathcal{H}_{dip}=\hbar\sqrt{\widetilde{N}}\sum_p\kla{\Omega_pb^\dagger a_p+H.c.}+
\sum_p\hbar(\omega_p-\omega_{p_0})a^\dagger_p a_p.
\end{equation}
The number of excited ions after their illumination is
\begin{equation}
\label{time_evolution_b}
\langle b^\dagger b\rangle_{t_M} =\sum_p\frac{|\Omega_p|^2}{\sum_q|\Omega_q|^2}\sin^2\kla{\sqrt{\sum_q|\Omega_q|^2}t_m}\langle a_p^\dagger a_p\rangle_0
\end{equation}
with the assumption of large Rabi frequencies $\sqrt{\widetilde{N}}\Omega_{p_0}\gg \omega_{p_0}$. Thus, the number of excited ions in state $\ket{b}$ is proportional to the number of phonons around the mode $\omega_{p_0}$. The standard deviation of the phonon number measurement is approximately
\begin{equation}
\sigma_m\approx \sqrt{\frac{\langle \kla{\Delta b^\dagger b}^2\rangle_{t_m}}{M}} \approx\frac{\langle \Delta b^\dagger b\rangle_{t_m}}{\sqrt{M}}\lesssim\frac{k_\text{B}T_\text{H}}{\hbar\omega_\text{rot}\sqrt{M}}.
\end{equation}

Now, we are presenting an example that satisfies the requirements of this measurement proposal. We consider $N=10^5$ singly charged $^9Be$ ions with ion spacing $L/N=2\mu m$ and average rotation frequency $\omega_\text{rot}=2\pi\times 1.2kHz$ ($e^2/4\pi\epsilon_0\approx 0.7/\kla{2N} mL^3/T^2$). The black hole region contains $(1-2\sigma)N=0.007N$ and the horizon region $2\gamma_1 N=0.004N$ ions. The angular velocity in the subsonic region is $v_\text{min}=2\pi\times 0.99/T$ and in the supersonic region $v_\text{max}=2\pi\times 2.4/T$. The Hawking temperature in this system is $k_\text{B}T_\text{H}/\hbar\approx 106/T\approx 2\pi\times 20kHz$. In the small time interval $\tau=0.05T$ of the black hole creation, the ions are normally traversing the angle $\Delta\theta\approx 2\pi\times 0.05$, which is large compared to the size of the black hole region. For the measurement we propose to illuminate $\widetilde{N}=200$ ions for the time $t_M=T/4$. In this case $M=100$ repetitions of the experiment are sufficient. 

\section{Conclusion}
In summary, we have discussed in this paper the details of a recent proposal to observe the Hawking effect with ions rotating on a ring \cite{Horstmann10}. We have described how to create an analog black hole spacetime in this system (see Sec. \ref{system}). The horizon emits Hawking radiation with a thermal spectrum (see Sec. \ref{simulations}). We have analyzed the emergence of correlations and entanglement between the inside and the outside of a black hole after its creation (see Sec. \ref{sec.correlations}). These correlations are a signature for the pair creation mechanism of Hawking radiation.

In this paper we have deepened our analysis of the emerging entanglement (see Sec. \ref{Entanglement Generation}). We cannot observe the creation of entanglement at too high initial temperatures. The generated cross-correlations, instead, remain present at arbitrarily high initial temperatures. Thus, we find the transition from the \emph{quantum} to the \emph{classical} Hawking effect (spontaneous versus stimulated emission).

Nevertheless, current technology allows to measure the \emph{quantum} Hawking effect in an experiment. We have presented a detailed discussion of realistic measurement techniques. It is possible to directly measure the cross-correlation signal (see Sec. \ref{Section Measurement}) or the emitted Hawking phonons (see Sec. \ref{Direct Measurement}).

To conclude, we expect ring traps to be extremely useful for future quantum simulations. They offer great opportunities, especially for studying translationally invariant systems.

\ack
We would like to thank T. Sch\"atz and D. Porras for many fruitful discussions and the German Excellence Initiative via the Nanosystems Initiative Munich and the German-Israeli Science foundation for financial support. S.F.'s research is supported by Anne McLaren fellowship. B.R. acknowledges the Israel Science Foundation grant 920/09 and the  European Commission (PICC). R.S. acknowledges support from the DFG (SFB-TR12 and SCHU~1557/1-3).

\begin{appendix}
\section{External Forces}
\label{system detail}
In this Appendix we present the detailed form of the external forces appearing in the general Hamiltonian \eqref{Hamiltonian general} and the harmonic Hamiltonian Eq. \eqref{Hamiltonian}. These forces are chosen to enforce the imposed equilibrium motion of the ions $\theta_i^0(t)$ in Eq. \eqref{definition g}.

The Coulomb force on the $i$th ion tangential to the ring is given by 
\begin{equation}
 F^\text{c}_i\kla{\theta_1(t),\dots,\theta_N(t)}=\frac{e^2}{4\pi\epsilon_0}\sum_{j\ne i} F^\text{c}\kla{\theta_i(t)-\theta_j(t)}
\end{equation}
with
\begin{equation}
F^\text{c}(\Delta \theta)=\pi^2\text{sign}(\sin( \Delta \theta/2)) \frac{\cos(\Delta\theta/2)}{L^2\sin(\Delta\theta/2)^2}, 
\end{equation}
where $\theta_i(t)$ gives the ion position at time $t$. The classical equations of motion are now $mL/(2\pi)\ddot{\theta}_i(t)=F^\text{c}_i(t)+F^\text{e}(\theta_i(t))$. We determine the local external force $F^\text{e}$ from this equation such that it guarantees the imposed equilibrium trajectories of the ions
\begin{eqnarray}
\label{eq.external force}
 F^\text{e}(\theta,t)&=&\klab{\frac{L}{2\pi T^2}}g''_{v_\text{min}}\kla{g^{-1}(\theta)}\nonumber\\
&&+\klab{\frac{L}{2\pi T}}2\frac{\partial g'_{v_\text{min}}}{\partial v}\kla{g^{-1}(\theta)}\cdot\frac{dv}{dt}(t)\nonumber\\
&&+\klab{\frac{L}{2\pi }}\frac{\partial^2 g_{v_\text{min}}}{\partial v^2}\kla{g^{-1}(\theta)}\cdot\left[\frac{dv}{dt}(t)\right]^2\nonumber\\
&&+\frac{\partial g_{v_\text{min}}}{\partial v}\kla{g^{-1}(\theta)}\cdot\frac{d^2v}{dt^2}(t)\nonumber\\
&&+\sum_{i=1}^{N-1} F^\text{c}\kla{\theta-g_{v_\text{min}}(g^{-1}(\theta)+\frac{i}{N})}.
\end{eqnarray}
This force is time-independent if the parameter $v_\text{min}$ is time-independent.

In harmonic approximation (see Eq. \eqref{Hamiltonian}) the Coulomb force and the external force are encoded in the force matrix $\mathcal{F}=\kla{f_{ij}}$
\begin{equation}
\label{eq.forcematrix}
 \kla{\frac{L}{2\pi}}^2f_{ij}(t)=f^\text{c}_{ij}(t)+\delta_{ij}f^\text{e}_i(t).
\end{equation}
The contribution from the Coulomb force is
\begin{gather}
 f^\text{c}_{ij}(t)=\frac{e^2}{4\pi\epsilon_0}\cdot\begin{cases}
           f^\text{c}\kla{\theta_i^0(t)-\theta_j^0(t)}& i\ne j\\
	   -\sum_{k\ne i}f^\text{c}\kla{\theta_i^0(t)-\theta_k^0(t)} &i=j
          \end{cases}
\end{gather}
with
\begin{equation}
 { f^\text{c}\kla{\Delta \theta}=\pi^3\left|\frac{1+\cos(\Delta\theta/2)^2}{\sin(\Delta\theta/2)^3}\right|}.
\end{equation}
The contribution from the diagonal external force is
\begin{eqnarray}
 f^\text{e}_i(t)&=&\frac{m}{g'\kla{\frac{i}{N}+\frac{t}{T}}} \left\{\klab{\frac{L}{2\pi T^2}}g'''_{v_\text{min}}\kla{\frac{i}{N}+\frac{t}{T}}\right.\nonumber\\
&&+2\klab{\frac{L}{2\pi T}}\frac{\partial g''_{v_\text{min}}}{\partial {v_\text{min}}}\kla{\frac{i}{N}+\frac{t}{T}}\cdot\frac{dv}{dt}(t)\nonumber\\
&&+\klab{\frac{L}{2\pi }}\frac{\partial^2 g'_{v_\text{min}}}{\partial {v_\text{min}}^2}\kla{\frac{i}{N}+\frac{t}{T}}\cdot\left[\frac{dv}{dt}(t)\right]^2\nonumber\\
&&\left.+\frac{\partial g'_{v_\text{min}}}{\partial {v_\text{min}}}\kla{\frac{i}{N}+\frac{t}{T}}\cdot\frac{d^2v}{dt^2}(t)\right\}\nonumber\\
&&+\frac{e^2}{4\pi\epsilon_0}\cdot\sum_{j\ne i}f^\text{c}\kla{\theta_i^0(t)-\theta_j^0(t)}\left[{1-\frac{f'\kla{\frac{j}{N}+t}}{f'\kla{\frac{i}{N}+t}}}\right].
\end{eqnarray}

\section{Velocity Profile}
\label{section velocity profile}
The classical equilibrium positions of the ions, thus their velocity profile, are imposed by the function $g$ as specified in Eq. \eqref{definition g}. We make the choice
\begin{equation}
\label{metricg}
\frac{g'(x)}{T}=\begin{cases}
       v_\text{min} & 0\le x\le \sigma-\gamma_1\\
	\beta+\alpha h\kla{\frac{x-\sigma}{\gamma_1}}	 & -\gamma_1<x-\sigma<\gamma_1\\ 
       v_\text{max} & \sigma+\gamma_1\le x\le 1-\sigma-\gamma_2\\
	\beta-\alpha h\kla{\frac{x-1+\sigma}{\gamma_2}} & -\gamma_2<x-\kla{1-\sigma}<\gamma_2\\
       v_\text{min} & 1-\sigma+\gamma_2\le x \le 1
      \end{cases}
\end{equation}
with $\alpha=\kla{v_\text{max}-v_\text{min}}/2$ and $\beta=\kla{v_\text{max}+v_\text{min}}/2$ and $h(s)=15/8s-5/4s^3+3/8s^5$. We use $g(0)=0$ to determine the equilibrium positions of the ions. In the supersonic region $\sigma v_\text{min}T \lesssim \theta\lesssim 2\pi-\sigma v_\text{min}T$ the constant angular ion velocity is $v_\text{max}$ (see Eq. \eqref{vmax}) and in the complementary subsonic region its constant value is $v\kla{\theta}=v_\text{min}$. Thus, the black hole horizon is located close to $\theta_\text{H}=\sigma v_\text{min}T$. This stepwise definition has the advantage that the velocity profile is flat inside and outside of the horizon, that the width of both regions can be adjusted, and that the width of the horizon regions can be adjusted. This flexibility is useful for the detection of correlation patterns in Sec. \ref{sec.correlations}. Also this profile must be sufficiently continuous to generate a physically allowed equilibrium motion.

In a part of this paper we dynamically create a black hole metric from a flat metric. We choose to reduce the velocity in the subsonic region from $2\pi/T$ at $t=0$, corresponding to homogeneously spaced ions, down to $v_\text{min}$ at $t\gg\tau$ according to
\begin{equation}
\label{Reduce vmin}
 v_\text{min}(t)=v_\text{min}+\kla{\frac{2\pi}{T}-v_\text{min}} \exp\left[-\kla{\frac{t}{\tau}}^2\right].
\end{equation}
We choose a Gaussian profile to guarantee $\frac{\partial{v}}{\partial{t}}(t=0)=0$. In this case, the velocity profile becomes (compare with Eq. \eqref{stationary velocity})
\begin{equation}
\label{vel}
 v(\theta,t)=\frac{g'_{v_\text{min}}\kla{g^{-1}\kla{\theta}}}{T}+\frac{\partial g_{v_\text{min}}}{\partial {v_\text{min}}}\kla{g^{-1}\kla{\theta}}
\cdot \frac{d{v_\text{min}}}{dt}(t).
\end{equation}

\section{System Dynamics and Equilibrium State}
\label{equations of motion}
The quasi-free quantum dynamics of the harmonic system \eqref{Hamiltonian} are governed by the classical linear equations of motion for the first (see Eq. \eqref{first moments}) and second moments (see Eq \eqref{Gamma}). The equations for the first moments can be written
\begin{equation}
\label{Newton}
\frac{\partial}{\partial t}\langle \hat{\xi}_i\rangle_t=\sum_j G_{ij}(t)\langle\hat{\xi}_j\rangle_t
\end{equation}
with the matrix $\mathcal{G}=\kla{G_{ij}}$
\begin{equation}
\mathcal{G}=\begin{pmatrix}
              0 & \kla{\frac{2\pi}{L}}^2\cdot\frac{1}{m} \\
	      \kla{\frac{L}{2\pi}}^2{\mathcal F} & 0 	
             \end{pmatrix}.
\end{equation}
The dynamics for the second moments are governed by the equation
\begin{equation}
\label{Gamma_dynamics}
\frac{\partial}{\partial t}\Gamma(t)=\mathcal{G}(t)\cdot\Gamma(t)+\Gamma(t)\cdot \mathcal{G}(t)^T.
\end{equation}
We determine the thermal state with temperature $T_0$ of homogeneously spaced ions at rest by a mode decomposition of the system. The Fourier transform $O$ diagonalizes the system \eqref{Hamiltonian}
\begin{equation}
\delta \widetilde{\theta}_k =\sum_{i=1}^N O_{ki}\delta\hat{\theta}_i, \hspace{2mm} \delta \widetilde{p}_k =\sum_{i=1}^N O_{ki}^T\delta \hat{p}_i
\end{equation}
to
\begin{equation}
\mathcal{H}=\sum_{k=1}^{N}\klab{ \frac{\delta \widetilde{p}_i^2}{2m}+\frac{m}{2}\omega_k^2{\delta\widetilde{\theta}}_k^2\kla{\frac{L}{2\pi}}^2}
\end{equation}
The mode frequencies are
\begin{equation}
 \omega_k^2\delta_{kl}=\kla{\frac{2\pi}{L}}^2\kla{OfO^T}_{kl}.
\end{equation}
According to the Bose-Einstein statistic each mode is on average occupied by
\begin{equation}
 \langle \hat{n}_k\rangle=\frac{1}{\exp\kla{\frac{\hbar\omega}{k_\text{B}T_0}}-1} 
\end{equation}
phonons. In equilibrium the first moments vanish $\langle \hat{\xi}_i\rangle=0$ due to the parity symmetry of the harmonic Hamiltonian \eqref{Hamiltonian}. The covariance matrix at temperature $T_0$ is given by 
\begin{gather}
 \langle \delta\hat\theta_i\delta\hat\theta_j\rangle=\kla{\frac{2\pi}{L}}^2 \frac{\hbar}{m}\sum_{k=1}^NO_{ik}^T\frac{\langle\hat{n}_k\rangle+\frac{1}{2} }{\omega_k}O_{kj},\nonumber\\
\langle \delta \hat{p}_i\delta\hat{p}_j\rangle= \hbar m \sum_{k=1}^NO_{ik} \omega_k\kla{\langle\hat{n}_k\rangle+\frac{1}{2} }O_{kj}^T,\nonumber\\
\langle \delta\hat{\theta}_i\delta \hat{p}_j\rangle=0.
\label{thermal state}
\end{gather}

\section{Stability Analysis}
\label{stability analysis}
\begin{figure}[t]
\begin{center}
\includegraphics[width=75mm,height=57mm,angle=0]{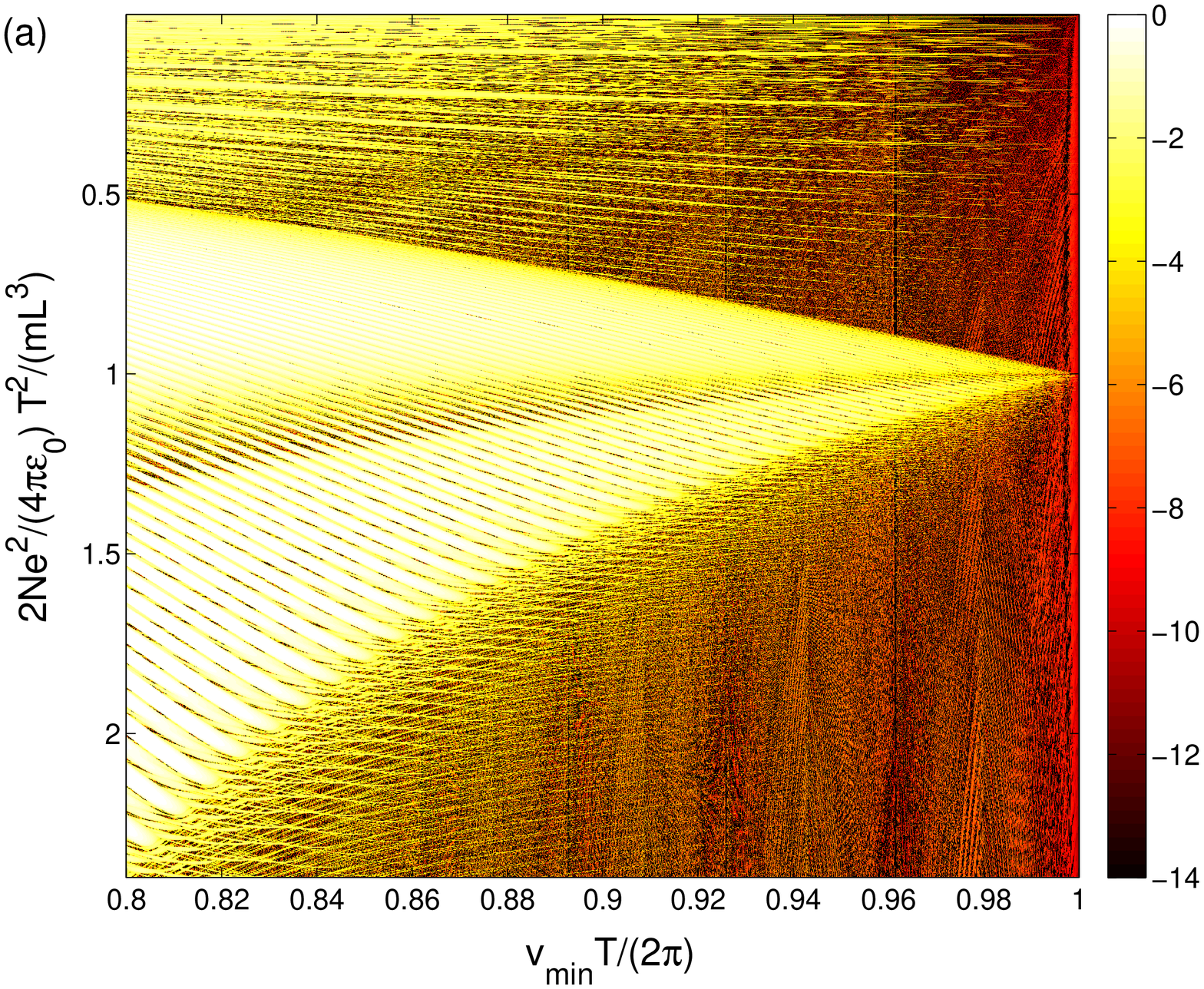}
\includegraphics[width=75mm,height=57mm,angle=0]{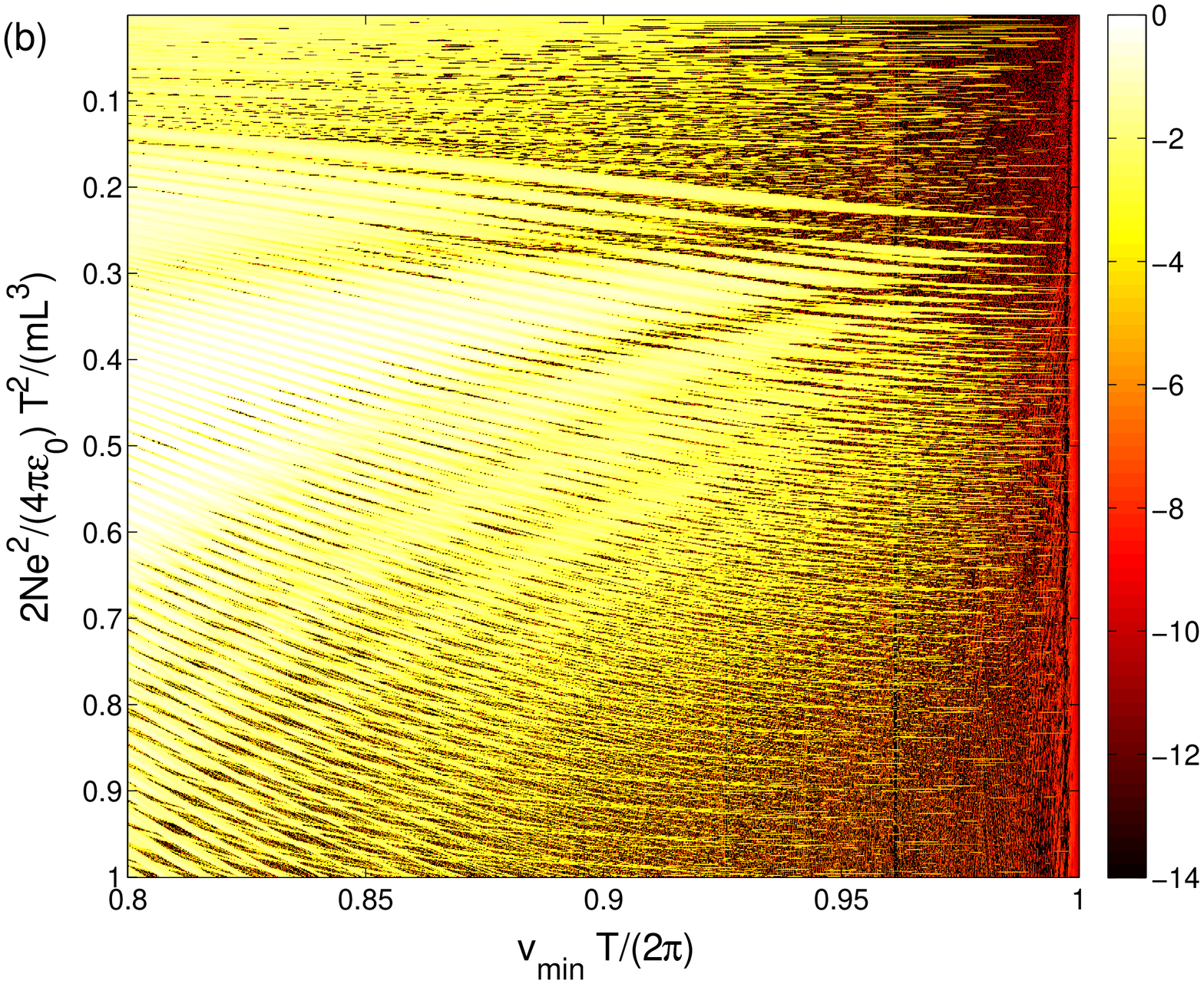}
\caption{$\log\kla{\mu-1}$ for $N=100$ ions, and $\sigma v_\text{min}T=2\pi\cdot 0.25$. In this stability diagram values of $\log\kla{\mu-1}$ close to zero represent stable systems. (a) Only nearest-neighbor interactions considered; (b) Full Coulomb interactions considered.}
\label{stability}
\end{center}
\end{figure}


\begin{figure}[t]
\begin{center}
\includegraphics[width=80mm,height=60mm,angle=0]{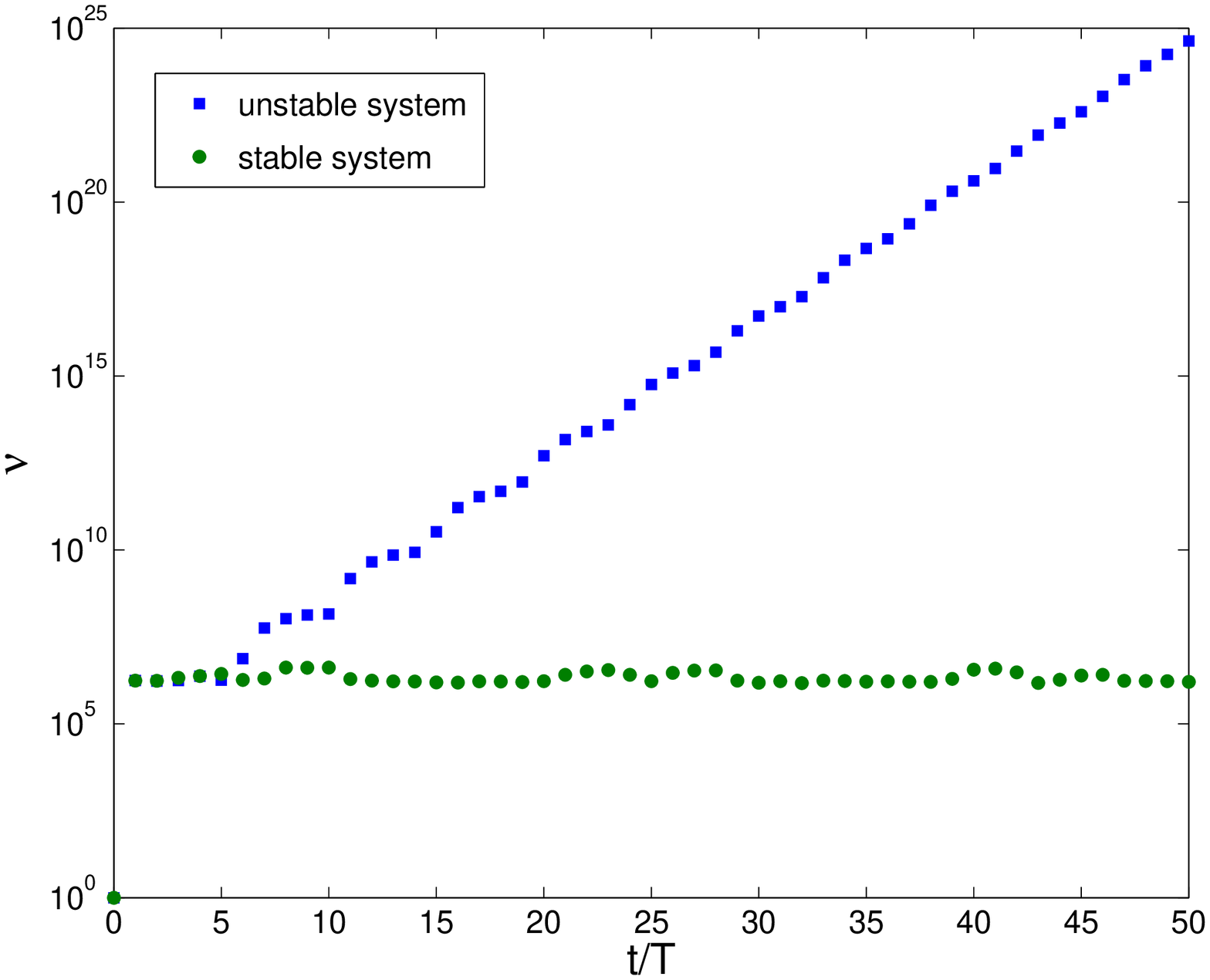}
\caption{Largest eigenvalue of $\mathbf U^\dagger(t)\mathbf U(t)$ at time steps t=nT. This quantity is an upper bound of observables. We consider $N=1000$ ions, $\sigma v_\text{min}T=2\pi\cdot 0.25$, and $\tau=0.05T$. We compare the stable system $e^2/4\pi\epsilon_0=\frac{0.2453}{2N}\frac{mL^3}{T^2}$ (green dots) with the unstable system $e^2/4\pi\epsilon_0=\frac{0.2446}{2N}\frac{mL^3}{T^2}$ (blue squares). Full Coulomb interactions are considered.}
\label{stabilitycomparison}
\end{center}
\end{figure}

In this Appendix we describe a stability analysis of our system. After a brief introduction to the method, we show results of the numerical simulations and put them into the context of previous works \cite{Cirac00}. Finally, we show that the observed instabilities are not important for our proposal.

The solution of explicit linear differential equations like Eq. \eqref{Newton} can be written
\begin{equation}
\label{evolution}
\langle \hat{\xi}_i\rangle_t=\sum_j U_{ij}(t)\langle\hat{\xi}_j\rangle_t
\end{equation}
with the monodromy matrix $\mathbf{U}(t)$ satisfying the initial condition $\mathbf{U}(0)=\mathbf{1}$ and the dynamics
\begin{equation}
 \partial_t\mathbf{U}(t)=\mathbf{G}(t)\cdot\mathbf{U}(t).
\end{equation}
If the system is periodic $\mathbf{G}(t+T)=\mathbf{G}(t)$, Floquet theorem states
\begin{equation}
 \mathbf{U}(t)=\mathbf{X}(t)\cdot e^{\mathbf{R}t}
\end{equation}
with the periodic matrix $\mathbf{X}(t+T)=\mathbf{X}(t)$ and the constant matrix $\mathbf{R}$ (see e.g., \cite{Floquet}). Therefore, the stability of the motion is determined by the eigenvalue of the matrix $\mathbf{U}(T)=\exp\kla{\mathbf{R}T}$ with the largest magnitude. We denote this magnitude by $\mu$, the dynamics are unstable for $\mu>1$. The monodromy matrix $\mathbf{U}$ does not only describe the first moments (see Eq. \eqref{evolution}), but also the evolution of the covariance matrix (see Eq. \eqref{Gamma_dynamics})
\begin{equation}
 \Gamma(t)=\mathbf{U}(t)\Gamma(0)\mathbf{U}^T(t)=\kla{\mathbf{U}(t)\otimes \mathbf{U}(t)}\Gamma(0).
\end{equation}
In our case the periodicity with period $T$ is not the highest symmetry. The system is also invariant under combined translations in time and space (see Eq. \eqref{higher symmetry}). Thus $\mathbf U(T)$ follows from $\mathbf U(T/N)$
\begin{equation}
\label{better}
\mathbf{U}(T)=\kla{\mathcal{T}\cdot\mathbf{U}(T/N)}^N,
\end{equation}
where $\mathcal T$ is the index translation matrix. 

The results of the numerical stability analysis are shown in Fig. \ref{stability}(a) for nearest-neighbor interactions and in Fig. \ref{stability}(b) for full Coulomb interactions as a function of $\overline{(cT/(2\pi))^2}=2Ne^2/(4\pi\epsilon_0)\cdot T^2/(mL^3)$ and $v_\text{min}T/(2\pi)$. For $\sigma v_\text{min}T=2\pi \times 0.25$, $\overline{c^2}$ is the mean of the squares of the two sound velocities (see Eq. \eqref{eq.sound velocity}) on the ion ring. The analysis shows a distinction between three parameter regions. If the system is supersonic on the whole ring (at small $\overline{c^2}$) the system is mostly stable with stripes of instabilities. If the system is subsonic on the whole ring (at large $\overline{c^2}$) the system is always stable. In the interesting intermediate regime a subsonic region coexists with a supersonic region on the ion ring, here the stability analysis is most complex. In this case the system is unstable apart from stripes of stability that bunch up in the central region which is most interesting for experiments. So we find that the appearance of instabilities is closely related to the presence of a black hole horizon. The structure of the instabilities is very similar to the result of the stability analysis in \cite{Cirac00}. Black hole laser instabilities \cite{Corley99} may contribute to these behavior. They occur in the presence of two horizon when particles bounce between the horizons and enhance themselves.

This stability analysis detects exponential instabilities. We show now that these exponential instabilities are not important for the proposed experiment. The increase of the experimental quantities is bounded by the maximal magnitude $\nu$ of the eigenvalues of $(\mathbf U^\dagger(T))^n(\mathbf U(T))^n$ for $n=1,2,\dots$. $\nu$ is shown in Fig. \ref{stabilitycomparison} for an exponentially stable and an exponentially unstable system. The exponential instability becomes dominant for $t>5T$. Therefore, it does not have a consequence for the proposed experiment performed during $t\lesssim T$. 
\end{appendix}

\end{document}